\documentclass[sigconf, nonacm]{acmart}
\usepackage{fontawesome}
\usepackage{bbm}
\usepackage{xcolor}
\usepackage{todonotes}
\usepackage{algorithm}
\usepackage{algpseudocode}
\usepackage{subcaption}
\usepackage{float}
\usepackage{cleveref}
\usepackage[normalem]{ulem}

\renewcommand\footnotetextcopyrightpermission[1]{} % Nocopyright block

\crefformat{equation}{#2#1(#3)}

\newcommand{\revised}[1]{\textcolor{black}{#1}}

\newcommand\vldbdoi{XX.XX/XXX.XX}

\newcommand\vldbavailabilityurl{URL_TO_YOUR_ARTIFACTS}
\newcommand\vldbpagestyle{plain}

\newcommand{\cp}{CP}
\newcommand{\stl}{STL}
\newcommand{\ml}{ML}

\newcommand{\lqo}{LQO}
\newcommand{\neo}{Neo}
\newcommand{\bao}{Bao}
\newcommand{\kepler}{Kepler}
\newcommand{\balsa}{Balsa}
\newcommand{\lero}{Lero}

\newcommand{\tpch}{TPC-H}
\newcommand{\job}{JOB}

\newcommand{\pdf}{PDF}

\begin{document}

\title{Conformal Prediction for Verifiable Learned Query Optimization}

\author{Hanwen Liu}
\orcid{}
\affiliation{%
  \institution{University of Southern California}
  \city{Los Angeles}
  % \state{California}
  \country{USA}
}
\email{hanwen_liu@usc.edu}

\author{Shashank Giridhara}
\authornote{Work done while at USC’s NexDIG group.}
\orcid{}
\affiliation{%
  \institution{Amazon Web Services}
  \city{Palo Alto}
  % \state{California}
  \country{USA}
  }
\email{smgiridh@amazon.com}
%\email{smgiridh@usc.edu}

\author{Ibrahim Sabek}
\orcid{}
\affiliation{%
  \institution{University of Southern California}
  \city{Los Angeles}
  % \state{California}
  \country{USA}
}
\email{sabek@usc.edu}

\begin{abstract}
Query optimization is critical in relational databases. Recently, numerous Learned Query Optimizers ({\lqo}s) have been proposed, demonstrating superior performance over traditional hand-crafted query optimizers after short training periods. However, the opacity and instability of machine learning models have limited their practical applications. To address this issue, we are the first to formulate the {\lqo} verification as a Conformal Prediction ({\cp}) problem. We first construct {\cp} model and obtain user-controlled \revised{bounded} ranges for the actual latency of {\lqo} plans before execution. Then, we introduce {\cp}-based runtime verification along with violation handling to ensure performance prior to execution. \revised{For both scenarios, we further extend our framework to handle distribution shifts in the dynamic environment using adaptive {\cp} approaches}. Finally, we present {\cp}-guided plan search, which uses actual latency upper bounds from {\cp} to heuristically guide query plan construction. We integrate our verification framework into three {\lqo}s (Balsa, Lero, and RTOS) and conducted evaluations on the {\job} and {\tpch} workloads. Experimental results demonstrate that our method is both accurate and efficient. Our {\cp}-based approaches achieve tight upper bounds, reliably detect and handle violations. \revised{Adaptive $\cp$ maintains accurate confidence levels even in the presence of distribution shifts}, and the {\cp}-guided plan search improves both query plan quality (up to 9.84x) and planning time, with a reduction of up to 74.4\% for a single query and 9.96\% across all test queries from trained {\lqo}s.

\end{abstract}

% \keywords{Database Systems, Query Optimizer, Learned Query Optimization, Conformal Prediction}

\maketitle

\pagestyle{\vldbpagestyle}
% \begingroup\small\noindent\raggedright\textbf{PVLDB Reference Format:}\\
% \vldbauthors. \vldbtitle. PVLDB, \vldbvolume(\vldbissue): \vldbpages, \vldbyear.\\
% \href{https://doi.org/\vldbdoi}{doi:\vldbdoi}
% \endgroup
\vspace{-24pt}
% \begingroup
% \renewcommand\thefootnote{}\footnote{\noindent
% This work is licensed under the Creative Commons BY-NC-ND 4.0 International License. Visit \url{https://creativecommons.org/licenses/by-nc-nd/4.0/} to view a copy of this license. For any use beyond those covered by this license, obtain permission by emailing \href{mailto:info@vldb.org}{info@vldb.org}. Copyright is held by the owner/author(s). Publication rights licensed to the VLDB Endowment. \\
% \raggedright Proceedings of the VLDB Endowment, Vol. \vldbvolume, No. \vldbissue\ %
% ISSN 2150-8097. \\
% \href{https://doi.org/\vldbdoi}{doi:\vldbdoi} \\
% }\addtocounter{footnote}{-1}
% \endgroup

\ifdefempty{\vldbavailabilityurl}{}{
\vspace{.3cm}
\begingroup\small\noindent\raggedright\textbf{PVLDB Artifact Availability:}\\
The source code, data, and/or other artifacts have been made available at \url{\vldbavailabilityurl}.
\endgroup
}

\section{Introduction}
\label{sec:introduction}

A query optimizer is a performance-critical component in every database system. It translates declarative user queries into efficient execution plans~\cite{selinger1979access,babcock2005towards}. %In particular, it navigates a broad search space of candidate plans for each query and assigns a score to each plan using statistics about the query and the underlying data. However, manually building an optimizer with good heuristics is tedious, especially when the database system's execution and storage engines evolve rapidly. 
%However, building a good optimizer today takes thousands of hours of engineering work to develop effective optimization heuristics and is a skill mastered by only a few experts. 
%As a result of this, 
There have been numerous efforts to learn query optimizers ({\lqo}s)(e.g.,~\cite{neo, bao, balsa,kepler}) to reduce the reliance on manual tuning and expert intervention, and ultimately lead to more intelligent and responsive database systems.
%Among the different {\ml} techniques, reinforcement learning ({\rl}) shows great promise in learning query optimizers (e.g.,~\cite{neo, bao, balsa}). %In an {\rl}-based optimizer, the database system itself acts as the environment, and the query planner acts as an agent that observes the environment state when an input query arrives and prepares an execution plan (i.e., an action) that minimizes the expected query latency (i.e., maximizing the reward). Using a feedback loop, the optimizer progressively improves its ability to generate effective query plans by learning from its successes and mistakes through trial and error across different queries in the user workload.
%%Within this context, the query planner is an agent that observes the current partial plan for a query (as an environment state) and adds operators to the partial plan (as an action) that minimizes the expected execution latency of the complete plan (i.e., maximizing the reward).
Unfortunately, {\lqo}s suffer \revised{three} main drawbacks. First, they can result in slow execution plans at the beginning of the learning process (sometimes orders of magnitude slower than the optimal plan~\cite{qoeval15}), where the probability of selecting disastrous plans is high. These disastrous plans at the beginning can slow the {\lqo}'s convergence to efficient query plans later. %Note that this is unlike other successful {\rl}-based applications, such as AlphaGo game~\cite{alphago16}, in which "bad" actions typically do not hinder the learning process (e.g., a bad player move can end the game quicker). 
Second, although {\lqo}s can outperform traditional optimizers on average, they often perform catastrophically (e.g., 100x query latency increase) in the tail cases, especially when the training data is sparse~\cite{neo}. \revised{Third, {\lqo}s are normally trained for specific workload. Their performance degrades significantly when distribution shifts exist in the query workloads and the underlying data~\cite{learnedce21, neo, negi2023robust}.}
% Third, {}the performance of these {\lqo}s degrades significantly when continuous distribution shifts exist in the query workloads and the underlying data~\cite{learnedce21, neo}.

Given these drawbacks, verifying that the {\lqo}'s generated plans satisfy the critical \textit{latency constraints} in real-life applications is crucial. Unfortunately, typical model checking techniques (e.g.,~\cite{formalisol16,formalatomic18}) that have been successfully investigated to verify the properties of other database components, such as transaction management and concurrency control, fail when the search space to be explored grows drastically as in query optimizers. In addition, statistical variations of these techniques (e.g.,~\cite{smcdb19}) do not perform verification during the runtime. Using these techniques, an {\lqo} might be verified to be constraint-compliant a priori. However, during runtime, we may observe certain query plans that violate the constraints due to the unknown changes in the execution environment. \revised{Additionally, these techniques should also be able to verify {\lqo}s operating in dynamic environments.
}

Meanwhile, Conformal Prediction ({\cp})~\cite{cpbook05,cpgentle23} has recently emerged as an efficient solution to perform runtime verification (e.g.,~\cite{rtfaulty10,rtltl11}) with \textit{formal guarantees} (e.g.,~\cite{cpstl23,prtstl22,neuralprt19,prtstl23}). In particular, {\cp} is a rigorous statistical tool to quantify the uncertainty of the {\ml} models' predictions while allowing users to specify the desired level of confidence in the quantification and being agnostic to the details of the {\ml} models. \revised{{\cp}-based runtime verification showed a great success in verifying many cyber-physical systems such as autonomous cars~\cite{cpstl23}, autonomous robots~\cite{prtrobots23}, and aircraft simulation~\cite{cpstl23,prtstl22}, among others. However, {\cp}-based runtime verification was never explored in the context of database systems before}. 

In this paper, we present the \underline{first} study of the {\lqo} verification problem using {\cp}. Specifically, we use {\cp} to solve the {\lqo} verification problem in two ways. First, we employ {\cp} to provide user-controlled \revised{bounded} ranges for the actual latency of constructed plans by {\lqo}s even before executing them (e.g., verifying that an {\lqo} plan for a specific query will never result in an execution time of more than 300 msec with a probability of at least 90\%). Second, we go further and explore the use of {\cp} to perform a runtime verification, with formal \revised{bounds}, that can early detect any performance constraint violation during the {\lqo}'s plan construction process based solely on the constructed partial plans so far and before the full plan is completed (e.g., with a user-defined confidence level of 95\%, we can detect at the second step of building a query plan by {\lqo} that the eventual complete plan will fail to satisfy a specific latency constraint). This will help in planning how to handle such violations during the plan construction time and before execution (e.g., falling back to a traditional query optimizer for re-planning). \revised{For both scenarios, we introduce an adaptive {\cp} framework to support {\lqo}s in static cases ({\lqo}s are trained and tested on the same workload) and in distribution shift cases (evaluating {\lqo}s on different workloads).
}Additionally, we propose a {\cp}-guided plan search algorithm that relies on upper bounds of the actual latency, instead of typical predicted costs by {\lqo}s, to generate more optimal query plans within shorter time frames. We also provide rigorous theoretical proofs of our approaches to ensure correctness and frameworks that facilitate the integration of our {\cp}-based verification approaches with {\lqo}s in real-world environments. 

% Our experimental results on the JOB~\cite{qoeval15} and TPC-H~\cite{tpch} workloads first confirm the correctness of applying {\cp} to {\lqo} by verifying empirical coverage. We then demonstrate the effectiveness of latency \revised{bounds} across multiple {\lqo}s, including Balsa~\cite{balsa}, Lero~\cite{zhu2023lero}, and RTOS~\cite{yu2020reinforcement}, all aligning with theoretical expectations. 

Our experimental results on the JOB~\cite{qoeval15} and TPC-H~\cite{tpch} workloads confirm the correctness of latency \revised{bounds} across multiple {\lqo}s, including Balsa~\cite{balsa}, Lero~\cite{zhu2023lero}, and RTOS~\cite{yu2020reinforcement}, all aligning with theoretical expectations. \revised{We then demonstrate the effectiveness of our adaptive {\cp} framework under distribution shift by evaluating it on workloads transitioning to CEB~\cite{Negi2021FlowLoss} and JOBLight-train~\cite{Kipf2019GroupBy}.} In runtime verification, we show that our {\cp}-based methods accurately detect violations, and our violation handling reduces overall execution latency by 12,030.1 ms across 7 violating queries. Using the {\cp}-guided algorithm, our approach improves plan quality in 33\% of queries from a moderately trained {\lqo}, achieving an additional 9.96\% reduction in overall planning latency across all test queries. For well-trained {\lqo}s, we observe better plan quality and faster query planning with our {\cp}-guided plan search algorithm. These comprehensive experiments substantiate the correctness and effectiveness of our {\cp}-based verification frameworks.

In summary, our \revised{novel} contributions are as follows: 
\begin{itemize}
    \item We are the first to formulate the Learned Query Optimizer ({\lqo}) verification as a Conformal Prediction ({\cp}) problem.
    \item We develop {\cp}-based latency \revised{bounds} for {\lqo}s, with formal proofs, to provide a user-defined confidence level a \revised{bounded} range for the actual latency of query plans.
    \item We design {\cp}-based runtime verification, with formal \revised{bounds}, which detect and address long-latency query plans even before completing the plan construction.
    \item \revised{We propose an Adaptive {\cp} framework for {\lqo}s which aids in handling distribution shifts, enhancing the robustness of the verification framework and making it suitable for real-world scenarios.}
    \item We introduce a \revised{generic} {\cp}-guided plan search algorithm that can enhance both the query plan quality and the planning time from a trained {\lqo}.
    \item Our experimental evaluation using the proposed {\cp}-based verification frameworks, across three {\lqo}s and \revised{four} workloads, demonstrates the correctness and effectiveness of our {\cp}-based frameworks for {\lqo}s.
\end{itemize}

We believe that our proposed {\cp}-based verification approaches hold promising potential for future applications across other learned components in database systems.

\vspace{-10pt}\section{Background}
\label{sec:background}

In this section, we first discuss the granularity levels of prediction decisions to be verified in learned query optimizers (Section~\ref{sec:background_lqo}). Then, we provide a brief introduction for the Conformal Prediction (Section~\ref{sec:background_cp}) and Signal Temporal Logic (Section~\ref{sec:background_stl}) tools that are used to build our verification framework and formally represent the performance constraints we verify {\lqo}s against, respectively.

\vspace{-6pt}\subsection{Granularity Levels of Decisions to be Verified in Learned Query Optimizers}
\label{sec:background_lqo}

\begin{figure}
    \centering
    \includegraphics[width=0.8\linewidth]{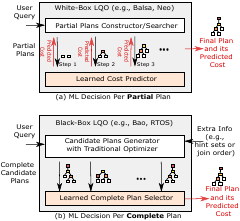}
    \vspace{-8pt}\caption{{\ml} Decisions in Learned Query Optimizers ({\lqo}s).}
    \label{fig:lqo_architecture}\vspace{-10pt}
\end{figure}

While Learned Query Optimizers ({\lqo}s) (e.g.,~\cite{neo, bao, balsa,kepler, zhu2023lero, yu2020reinforcement}) can improve the performance over traditional optimizers by adapting to complex queries and data distributions, their reliance on {\ml} models to take decisions introduces variability and potential unpredictability in performance. Therefore, verifying {\lqo}s against user-defined performance constraints is crucial to ensure that generated plans meet specific efficiency and reliability standards (e.g., the execution time of a specific query should be $\leq$ 100ms). Broadly, {\lqo}s fall into three categories based on how {\ml} is used. The first category uses {\ml} to improve specific components of the optimizer (e.g., cardinality estimator~\cite{learnedce21,deep_card_est2,neurocard} and cost estimator~\cite{learned_cost_estimator_1,learned_cost_estimator_2}). The second category uses {\ml} to construct the query plan from scratch, replacing the traditional optimizer (e.g.,~\cite{balsa, neo}). The third category uses {\ml} to steer the traditional optimizer in constructing better candidate plans and/or in selecting among them (e.g.,~\cite{bao,zhu2023lero,yu2020reinforcement}). In this paper, we focus on verifying the {\ml} decisions made by {\lqo}s in the second and third categories only, where {\ml} is involved in constructing the query plan itself. However, the granularity level of these decisions differs between these two categories. Figure~\ref{fig:lqo_architecture} shows a high-level overview of these two {\lqo} categories, highlighting their {\ml} decisions in red. In the second category, fine-grained prediction decisions are performed to construct the query plan step-by-step and predict the associated cost at each step\footnote{In this paper, we assume that the cost of a plan is indicative of its actual latency, where a higher cost corresponds to longer latency.}. For instance, {\balsa}~\cite{balsa} uses a learned value model to construct the optimized plan operator-by-operator and predict the intermediate cost for the final plan construction at each operator. We refer to the second category as \textit{white-box} {\lqo}s because we rely on these fine-grained prediction decisions during the verification process. In contrast, in the third category, learned models neither perform step-by-step plan construction nor intermediate cost predictions. Instead, these models are used to select the best plan from a set of candidate plans, either by predicting the high-level cost for each candidate~\cite{bao} or by assigning a relative rank to all candidates~\cite{yu2020reinforcement}. These candidate plans are typically constructed by a traditional optimizer and based on auxiliary information, such as join orders~\cite{yu2020reinforcement} and hint sets~\cite{bao}. Therefore, in this category, the selection decisions are mainly only on the level of the whole plan and its high-level associated cost, if available. We refer to the third category as \textit{black-box} {\lqo}s because we only access coarse-grained plan-level decisions (i.e., no partial-plan-level predictions) during the verification process.

\vspace{-6pt}\subsection{\revised{Standard Conformal Prediction ({\cp})}}
\label{sec:background_cp}

We build our {\lqo} verification framework, as shown later, based on Conformal Prediction ({\cp})~\cite{cpbook05,cpgentle23}, a rigorous statistical tool that efficiently quantifies the uncertainty of the {\ml} models' predictions. {\cp} enables users to specify the desired level of confidence in the quantification while being agnostic to the details of the {\ml} models. To introduce {\cp}, assume that \(R^{(0)}, R^{(1)}, \dots, R^{(K)}\) are \(K + 1\) independent and identically distributed (i.i.d) random variables, where each variable \(R^{(i)}\) for \(i \in \{0, \dots, K\}\) is an estimate of the prediction error between the true output \(y^{(i)}\), i.e., ground truth, for input \(x^{(i)}\) and the predicted value of this output $\eta(x^{(i)})$ by the {\ml} predictor~\(\eta\). Formally, this error can be expressed as:
\[
R^{(i)} := \|y^{(i)} - \eta(x^{(i)})\|,
\]
where $\lVert \cdot \rVert$ denoting the absolute value. \(R^{(i)}\) is commonly referred to as the \textit{non-conformity score}, where a small score suggests a strong predictive model and a large score indicates poorer performance (i.e., less accurate predictions). 

Now, assuming that $R^{(0)}$ belongs to test data and $R^{(1)}, \ldots, R^{(K)}$ are calibration data, the objective of {\cp} is to quantify the uncertainty of \(R^{(0)}\) using \(R^{(1)}, \dots, R^{(K)}\). Specifically, for a user-defined uncertainty probability $\delta \in [0, 1]$ (i.e., $1-\delta$ is a confidence level), {\cp} aims to compute an upper bound $C(R^{(1)}, \dots, R^{(K)})$ for the prediction error $R^{(0)}$ such that:
\[
\text{Prob}(R^{(0)} \leq C(R^{(1)}, \dots, R^{(K)})) \geq 1 - \delta \tag{1}\label{2.2-1}
\]
\revised{This upper bound \(C(R^{(1)}, \dots, R^{(K)})\) can be efficiently determined by computing the $(1-\delta)$th quantile of the empirical distribution of $R^{(1)}, \ldots, R^{(K)}$ and $\infty$, assuming training, calibration, and testing data originate from the same underlying distribution (i.e., the scores $R^{(0)}, R^{(1)}, \ldots, R^{(K)}$ are exchangeable)~\cite{cpgentle23}. Although this assumption aligns with the data and workload scenarios used in most state-of-the-art workload-aware {\lqo}s (e.g.,~\cite{balsa, bao, neo}), we extend our {\lqo} verification framework to support adaptive {\cp} for distribution shifts~\cite{zhao2024robust} as shown later in Section~\ref{sec:cp_for_lqo_dynamic}.} For simplicity, we will refer to the upper bound \(C(R^{(1)}, \dots, R^{(K)})\) as \(C\) in the rest of the paper. Note that {\cp} guarantees marginal coverage, which is not conditional on the calibration data~\cite{cpgentle23}.

%Additionally, our verification framework is built on the standard {\cp}~\cite{cpbook05,cpgentle23} variation that assumes training, calibration, and testing data originate from the same underlying distribution (i.e., the scores $R^{(0)}, R^{(1)}, \ldots, R^{(K)}$ are exchangeable~\cite{cpgentle23}). This aligns with the data and workload assumptions used in most state-of-the-art workload-aware {\lqo}s (e.g.,~\cite{balsa, bao, neo}). We will extend our verification framework to support {\cp} approaches for distribution shifts~\cite{cpcovshift19, zhao2024robust} in future work.  

%In both cases, at each step of query plan generation, a predicted cost is computed. This predicted cost quantifies the resources required based on a cost model. Once the plan is generated, it is executed by the query optimizer, and the actual runtime for each step is obtained. These predicted costs and actual runtimes are key performance indicators of the database and form the core of the research presented in this paper.

% \revised{\input{sections/hanwen_distribution_shift}}

\vspace{-6pt}\subsection{Formal Representation of Performance Constraints to be Verified with {\cp}} 
\label{sec:background_stl}
  
To formally represent the desired performance constraints to verify against {\lqo}s, we employ Signal Temporal Logic ({\stl})~\cite{robuststl10}, a {\cp}-compliant formal logical language for verification. {\stl} was originally introduced to verify the properties of time series data (e.g., signals), especially in the context of cyber-physical systems~\cite{stlintro04}. \revised{STL can also handle non-traditional time-series data where sequence or order matters. An} {\stl} specification $\phi$ is recursively defined as $\phi := True \mid \mu \mid \neg \phi \mid \phi \wedge \psi  \mid \mathbf{G}_{[a,b]} \phi$, where $\psi$ is an {\stl} formula. \revised{$\neg$ and $\wedge$ are the \textit{not} and \textit{conjunction} operators, respectively}. The \textit{always} operator $\mathbf{G}_{[a,b]} \phi$ encodes that $\phi$ has to be always true for the entire duration or steps between $a$ and $b$. $\mu$ is a predicate to check whether the semantics of the specification $\phi$ are achieved or not, i.e., \(\mu: \mathbb{R}^n \to \{\text{True}, \text{False}\}\). \revised{For instance, we can define an operator $\mathbf{G}_{[0, N-1]} \phi$ to check whether the query plan generated by a {\lqo} will \textit{always} have a latency less than $750$ msec at each of its $N$ execution steps (i.e., partial plans)}. \revised{In this case, $x := (x_0, x_1, \ldots, x_{N-1})$ will represent the partial plan latencies at steps~$0, 1, \ldots, {N-1}$ and the condition \( x_\tau < 750 \) forms the semantics of the specification $\phi$ that needs to be checked at each step $\tau$}.

Moreover, we can use \textit{robust} semantics $\rho^{\phi} (x)$, as in~\cite{robuststl09,robuststl10}, to extend the binary evaluation of {\stl} satisfaction (i.e., $\mu(x)$) by providing a quantitative measure of the degree to which this satisfaction is achieved. Unlike traditional binary satisfaction, robust semantics $\rho^{\phi} (x)$ produces a real-valued metric: positive values indicate that the specification $\phi$ is satisfied, with the magnitude representing the strength of satisfaction, whereas negative values denote a \textit{violation}, with the magnitude reflecting the severity of the violation. \revised{For example, considering the previously discussed specification $\phi$ with condition  \(x_\tau < 750 \), the robust satisfaction \( \rho^\phi(x) \) can be defined to provide a quantitative measure of how robustly all latencies $x$ satisfy this condition by calculating \( (750 - x_\tau) \) for each $x_\tau \in x$. In this case, $x_\tau = 100$ exhibits stronger robustness in satisfying  $\phi$ than $x_\tau = 600$, whereas $x_\tau =  800$ results in a \textit{violation}.} More details about robust {\stl} semantics are in~\cite{robuststl09,robuststl10}.

\vspace{-8pt}\revised{% \section{Adaptive Conformal Prediction for Handling Out-of-Distribution Queries}

}
\vspace{-8pt}%\section{{\cp}-based Latency Guarantees and Runtime %Verification for {\lqo} Plans }
%\label{sec:problem_formulation}

%In this section, we first describe how {\cp} is used to obtain guaranteed latency ranges for the plans generated by {\lqo}s before executing them (Section~\ref{sec:cp_for_lqo}). Then, we show how {\cp} can support runtime verification during the plan construction process, providing guarantees that the evolving plans satisfy user-defined performance constraints (Section~\ref{sec:cp_based_runtime_verification}).

\section{{\cp}-based Latency \revised{Bounds} for {\lqo}s}
%\subsection{{\lqo}s Prediction Error Analysis using {\cp}}
\label{sec:cp_for_lqo}

%You give an example on the lower value we have for C, the tighter bounds we can get …. 

As mentioned in Section~\ref{sec:background_lqo}, we focus on two categories 
of {\lqo}s: \textit{white-box} and \textit{black-box}, both of which use learned models to construct the query plan itself. In \textit{white-box} {\lqo}s (e.g.,~\cite{balsa, neo}), the learned model builds the query plan incrementally, constructing one partial plan at a time based on a predicted cost (Figure~\ref{fig:lqo_architecture}~(a)). %The predicted cost is designed to reflect the real execution performance of the whole plan when the corresponding sub-plan is used as a step in it: lower costs should correlate with lower execution latency. 
%These partial plans are then assembled into a complete query plan, executed, and its actual latency is used to fine-tune the learned model further. 
Here, we employ {\cp} to obtain \textit{user-controlled \revised{bounded} ranges} for the \textit{actual latency} (not the predicted cost) of these constructed partial plans \textit{before executing them}. For example, given a partial plan $s$ and a user-defined confidence level of 90\%, we can determine a latency range $[l^s_{min}, l^s_{max}]$ that the latency $l^s$ of $s$ will fall within with \textit{at least} 90\% probability, where $l^s_{min}$ and $l^s_{max}$ represent the lower and upper latency bounds, respectively. The intuition is to leverage {\cp} to gain insights into the relationship between predicted costs and actual latencies of partial plans from the {\lqo}'s calibration query workloads, and then use these insights to obtain latency ranges for testing queries. 
%Recall that we assume both calibration and testing queries share the same underlying distribution (check Section~\ref{sec:background_cp}). 
%Therefore, applying calibration-based insights to testing queries is valid.
Similarly in \textit{black-box} {\lqo}s (e.g.,~\cite{bao, yu2020reinforcement}), we use {\cp} to provide such user-controlled \revised{bounded} ranges, yet for the end-to-end latencies of complete plans rather than partial ones. This is because \textit{black-box} {\lqo}s rely on learned models solely to select the best plan among complete candidates (Figure~\ref{fig:lqo_architecture}~(b)). %~\footnote{In our {\cp} framework, we consider \textit{black-box} {\lqo}s as a special case of \textit{white-box} {\lqo}s, where the learned model only makes a single, high-level prediction for a large sub-plan at the root.}. 

\textbf{Latency-Cost Non-conformity Score.} A critical step in applying {\cp} is defining the non-conformity score $R$ (check Section~\ref{sec:background_cp}), as it quantifies the deviation between the predicted and actual outcomes. In the {\lqo} context, we focus on how the actual latency of a plan, whether partial or complete, deviates from its predicted cost~\footnote{Note that, during calibration, we can obtain the actual latency of any partial or complete plan straightforwardly with tools like EXPLAIN ANALYZE in PostgreSQL~\cite{postgres_explain_analyze}.}. Following the {\cp} notation, we formally define a \textit{latency-cost non-conformity score} $R^{(i)}$ for the plan at step $\tau$ in a query $q_{j}$ to be:
\[
R^{(i)} := \lVert t_{\tau}^{(j)} - \hat{c}_{\tau}^{(j)} \rVert \tag{2}\label{3.1-1}
\]
where $t_{\tau}^{(j)}$ is the actual latency of this plan and $\hat{c}_{\tau}^{(j)}$ is its predicted cost. Note that $R^{(i)}$ represents a score for a calibration plan (i.e., $R^{(i)} \in \{R^{(1)}, \ldots, R^{(K)}\}$) when $q_j$ belongs to the calibration workload $\mathcal{Q}^{Cal}$ and represents a score for a testing plan $R^{(0)}$ when $q_j$ belongs to the testing workload $\mathcal{Q}^{Tst}$. 

%(check Section~\ref{sec:background_cp})

\revised{In the following, we introduce our approach for using {\cp} to obtain the \revised{bounded} latency ranges when the calibration and testing distributions are similar, i.e., static case, (Section~\ref{sec:cp_for_lqo_static}), and then we extend it to handle distribution shifts in the testing distribution, i.e., distribution shift case (Section~\ref{sec:cp_for_lqo_dynamic}). Finally, we detail our proposed verification framework (Section~\ref{sec:cp_latency_guarantee_framework}).}

%Note that Equation~\ref{3.1-1} can be used to calculate the non-conformity scores of sub-plans in any query $q_j$ belonging to the calibration $\mathcal{Q}^{Cal}$ or testing $\mathcal{Q}^{Tst}$ query workloads.

%Note that Equation~\ref{3.1-1} is used to calculate the calibration $\{R^{(1)}, \ldots, R^{(K)} \}$ and testing $R^{(0)}$ scores, corresponding to the sub-plans in the calibration $\mathcal{Q}^{Cal}$ and  testing $\mathcal{Q}^{Tst}$ query workloads, respectively, i.e., $R^{(i)} \in \{R^{(0)}, R^{(1)}, \ldots, R^{(K)}\}$.

%Formally, assume that $t_{\tau}^{(j)}$ is the actual latency of a sub-plan at step $\tau$ in query $q_{j}$ and $\hat{c}_{\tau}^{(j)}$ is the predicted cost of this sub-plan, then we define a latency-cost non-conformity score $R^{(i)}$ as follows:

%\[
%R^{(i)} := \lVert t_{\tau}^{(j)} - \hat{c}_{\tau}^{(j)} \rVert \tag{2}\label{3.1-1}
%\]

%where $R^{(i)} \in \{R^{(0)}, R^{(1)}, \ldots, R^{(K)}\}$ such that $\{R^{(1)}, \ldots, R^{(K)} \}$ is a set of $K$ non-conformity scores defined over all sub-plans of $m$ queries in the calibration workload $\mathcal{Q}^{Cal}$ (i.e., if $q_j \in \mathcal{Q}^{Cal}$), and $R^{(0)}$ is a non-conformity score defined over a sub-plan of a query in the testing workload $\mathcal{Q}^{Tst}$ (i.e., if $q_j \in \mathcal{Q}^{Tst}$).

%%%TODO: maybe we should say R = |t-funct(c)| and func is for normalization
%%%TODO: Put an example here for these  different sub-plans from queries and 
\subsection{\revised{Latency Bounds in Static Cases}}
\label{sec:cp_for_lqo_static}

% Using equations \ref{2.2-1} and~\ref{3.1-1}, we can directly derive an upper bound~\(C\) on the latency-cost non-conformity scores of any plan, whether partial or complete, in a testing query such that:
\revised{Using equations \ref{2.2-1} and~\ref{3.1-1}, we can directly derive an upper bound~\(C\) on the latency of any plan, whether partial or complete, in a testing query as the $(1-\delta$)th quantile of the latency-cost non-conformity scores such that:}
\[
P(\lVert t_{\tau}^{(j)} - \hat{c}_{\tau}^{(j)} \rVert \leq C) \geq 1 - \delta \label{4.3-1}\tag{3}
\]
By reformulating Equation~\ref{4.3-1}, we can compute a range for the actual latency $t_{\tau}^{(j)}$ of this plan, with confidence \(1 - \delta\), based on its predicted cost $\hat{c}_{\tau}^{(j)}$ and the upper bound \(C\) as follows:
\[
P(\hat{c}_{\tau}^{(j)} - C \leq t_{\tau}^{(j)} \leq \hat{c}_{\tau}^{(j)} + C) \geq 1 - \delta \label{4.3-2}\tag{4}
\] 
This allows us to estimate a \textit{\revised{bounded}} range for the actual latency even prior to executing the plan. However, the tightness of this range primarily depends on the upper bound~\(C\), which itself is influenced by the number of calibration plans used to establish it. Therefore, determining the sufficient number of calibration plans to construct a valid upper bound~\(C\) is crucial. Here, we derive a lower bound on this number: \\

%Even though the number of calibration plan, \(K\), does not directly influence the upper bound \(C\). However, ensuring a valid upper bound \(C\) (rather than \(\infty\)) requires a minimal number of calibration data points.

%as the $(1-\delta)$th quantile of the empirical distribution of the $K$ non-conformity scores of calibration sub-plans $R^{(1)}, \ldots, R^{(K)}$ and $\infty$, assuming the availability of independent $\mathcal{Q}^{Tra}$ and $\mathcal{Q}^{Cal}$ query workloads (and their sub-plans latencies) for training and calibrating the {\lqo}, respectively

\vspace{-8pt}

\textbf{Lemma~1} \textbf{(Lower Bound on Required Calibration Plans)}. \textit{Let the latency-cost non-conformity scores of a testing plan $R^{(0)}$ and $K$ calibration plans $R^{(1)}, \ldots, R^{(K)}$ be exchangeable and realizing i.i.d random variables, $\delta \in [0, 1]$ be a user-defined uncertainty probability, and $C$ be an upper bound on the score $R^{(0)}$ of the testing plan, calculated at a confidence level of $1-\delta$. Then, the lower bound on the number of calibration plans, i.e., $K$, to calculate $C$ is $\frac{1 - \delta}{\delta}$.}\\

\vspace{-8pt}

\textbf{Proof.} If the scores $R^{(0)}, R^{(1)}, \ldots, R^{(K)}$ are exchangeable (i.e., independent of their order and are drawn from the same distribution), then the joint distribution of these scores remains unchanged~\cite{cpgentle23}. This means that the rank of any score, including $R^{(0)}$, is \textit{uniformly} distributed on the ranks $\{1, \ldots, K+1\}$. As a result, we can estimate the probability of the $R^{(0)}$'s rank in this uniform distribution using the $1-\delta$ quantile as follows:
\[
\text{Prob} (\text{Rank of } R^{(0)} \leq \lceil (K + 1)(1 - \delta) \rceil) \geq 1 - \delta
\]
where \(\lceil \cdot \rceil\) denoting the ceiling function. However, according to~\cite{cpgentle23}, if \(\lceil (K+1)(1-\delta) \rceil > K\), then the upper bound \(C\) becomes trivial and uninformative, yielding \(C = \infty\). Therefore, to ensure that \(C\) is nontrivial, we need the following condition: 
\[\lceil (K+1)(1-\delta) \rceil \leq K\]

From this, we can easily get \(K \geq \frac{1 - \delta}{\delta}\), which means the lower bound on the number of calibration plans should be $\frac{1 - \delta}{\delta}$.

%%% TODO: show some examples with delta 0.1 or 0.05

\subsection{\revised{Latency Bounds in Distribution Shift Cases}}
\label{sec:cp_for_lqo_dynamic}
\revised{In the preceding section, we assumed that the test data $\{R^{(0)}\}$ and the calibration data $\{R^{(1)}, \dots, R^{(K)}\}$ are drawn from the same underlying distribution. However, this assumption does not hold in workload drift scenarios, i.e., new or evolving workloads, that are common in database applications~\cite{negi2023robust,wu2023addingdomainknowledgequerydriven,10.1145/3639293}. For instance, slight changes in query patterns (e.g., filters on new columns), can violate the exchangeability assumption of $R^{(0)}, R^{(1)}, \ldots, R^{(K)}$ (see Section~\ref{sec:background_cp}), leading to an invalid upper bound $C$. To address this, we adopt an adaptive {\cp} variation, inspired by~\cite{Cauchois01102024}, which dynamically adjusts the upper bound to be $\tilde{C}$ based on the distribution shift in the testing workload only, assuming that this shift can be empirically estimated. This approach ensures that the newly calculated bounded latency range, based on $\tilde{C}$, preserves the user-specified confidence level \(1 - \delta\), even in the presence of distribution shifts.}

\revised{Specifically, let $\mathcal{D}$ represent the distribution of the testing workload (i.e., $R^{(0)} \sim \mathcal{D}$) and $\mathcal{D}_{0}$ represent the distribution of the calibration workload (i.e., $R^{(1)}, \dots, R^{(K)} \sim \mathcal{D}_{0}$). We can rigorously quantify the deviation between the calibration and test distributions using the total variation distance $TV(\mathcal{D}, \mathcal{D}_0)= \frac{1}{2} \int_x | P(x)-Q(x)| dx,$ where $P(x)$ and $Q(x)$ denote the probability density functions ({\pdf}s) of $\mathcal{D}$ and $\mathcal{D}_{0}$, respectively~\cite{devroye2013probabilistic}. To realize this in our {\lqo} context, we empirically estimate these PDFs of latency-cost non-conformity scores using kernel density estimators (KDEs) as Gaussian kernels. According to ~\cite{Cauchois01102024, zhao2024robust}, we can compute an adjusted uncertainty probability $\tilde{\delta}$ to account for the distribution shift from $\mathcal{D}_{0}$ to $\mathcal{D}$ as follows:
\[
    \tilde{\delta} = 1 - g^{-1}\left(g\left( \left( 1 + \frac{1}{K} \right) g^{-1}(1 - \delta) \right)\right) \tag{5}\label{equ:tilde_delta_complex}
\]}
\revised{
where \(\delta\) is the original user-specified uncertainty probability, \(K\) is the number of calibration plans, and \( g(\beta) = \max(0, \beta - \epsilon) \) and its inverse \( g^{-1}(\beta) = \min(1, \beta + \epsilon) \) are two functions calculated based on the allowable distribution shift $\epsilon$, which must be set to a value greater than or equal to $TV(\mathcal{D}, \mathcal{D}_0)$.
}
\revised{Then, similar to Equation~\ref{4.3-2}, the new latency bounds are calculated as:  
\[
P(\hat{c}_{\tau}^{(j)} - \tilde{C} \leq t_{\tau}^{(j)} \leq \hat{c}_{\tau}^{(j)} + \tilde{C}) \geq 1 - \delta \label{4.3-3}\tag{6}
\]}
\revised{
where $\tilde{C}$ is the \((1-\tilde{\delta})\)th quantile of the latency-cost non-conformity scores from the original calibration workload \(\mathcal{Q}^{Cal}\sim \mathcal{D}_{0}\).
}

\subsection{Framework Overview}
\label{sec:cp_latency_guarantee_framework}

\begin{figure}
    \centering
    \includegraphics[width=\linewidth]{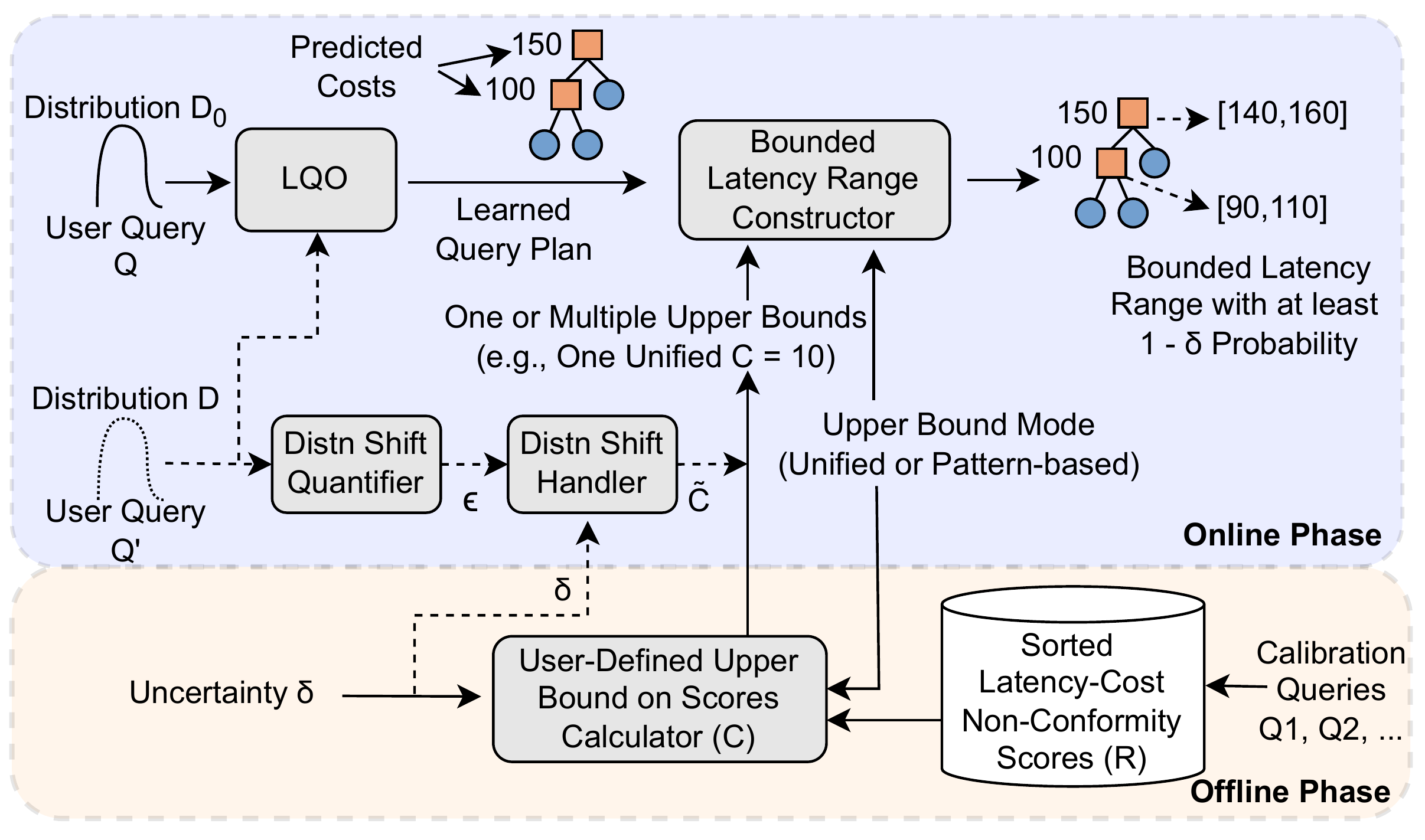}
    \vspace{-8pt}\caption{{\cp}-based \revised{Bounded} Latency Range Framework.}
    \label{fig:cp4lqo_latency_guarantee}\vspace{-10pt}
\end{figure}

% \begin{figure}
%     \centering
%     \includegraphics[width=\linewidth]{figures/cp4lqo_latency_guarantee.eps}
%     \caption{Original{\cp}-based \revised{Bounded} Latency Range Framework.\textbf{Just for comparison}}
%     \label{fig:cp4lqo_latency_guarantee}
% \end{figure}

Figure~\ref{fig:cp4lqo_latency_guarantee} gives an overview of our {\cp}-based framework to provide \revised{bounded} latency ranges before execution. 

\noindent\textbf{Offline Phase.} After training the {\lqo}, we first construct a set of latency-cost non-conformity scores using all plans - whether partial or complete - from the calibration query workload $\mathcal{Q}^{Cal}$. For each plan, we collect its predicted cost during the {\lqo}'s planning phase and its actual latency from execution. \revised{These scores are then sorted in ascending order and stored to be used along with the user-specified uncertainty probability \(\delta\) to compute any upper bound, whether $C$ in the static case or $\tilde{C}$ in the distribution shift case}.

\noindent \textbf{Online Phase.} The user first submits a testing query to the trained {\lqo}, which generates a query plan with predicted costs (either per partial plan for white-box {\lqo}s or a single cost for the entire plan in black-box  {\lqo}). 
%The user also specifies an uncertainty probability \(\delta\), which is used to construct the upper bound $C$ for generating a \revised{bounded} latency range. 
\revised{In case there is a distribution shift in the testing queries from \(\mathcal{D}_0\) to \(\mathcal{D}\), queries are also sent (represented by a dashed line) to a distribution shift quantifier to determine the allowable distribution shift \(\epsilon\) (check Section~\ref{sec:cp_for_lqo_dynamic}). This value, along with the user-defined parameter \(\delta\), is then used to construct the adjusted upper bound \(\tilde{C}\)}. Hereafter, we will use $C$ to denote the upper bound for both the static and distribution shift cases, as they are applied identically in subsequent steps. We support two modes for calculating the upper bound, namely \textit{Unified} and \textit{Pattern-based}, depending on the desired granularity level. In the \textit{Unified} mode, non-conformity scores from all partial and complete plans are treated equally to construct a single upper bound value for $C$, applicable to both partial and complete plans of the testing query~\footnote{Note that unified upper bounds can be calculated for white-box and black-box {\lqo}s}. In the \textit{Pattern-based} mode,  we account for the internal structure of partial plans by setting a unique $C$ value for each parent-child pattern. This $C$ value is applied only when that pattern appears in the testing query. Note that pattern-based upper bounds are available only for white-box {\lqo}s and are effective if we have sufficient calibration scores for each pattern (i.e., meeting the lower bound $K$ in Lemma~1 for each pattern). Otherwise, the unified upper bound is preferable. Algorithm~\ref{alg:cp_upper_bound} illustrates how to construct the two types of upper bounds given a specific user-defined uncertainty probability $\delta$. Once the upper bound(s) construction is done, the query plan along with the upper bound(s) are passed to the \revised{bounded} latency range constructor to obtain the \revised{bounded} ranges as in Equation~\ref{4.3-2}.

\begin{figure}
    \centering
    \includegraphics[width=\linewidth]{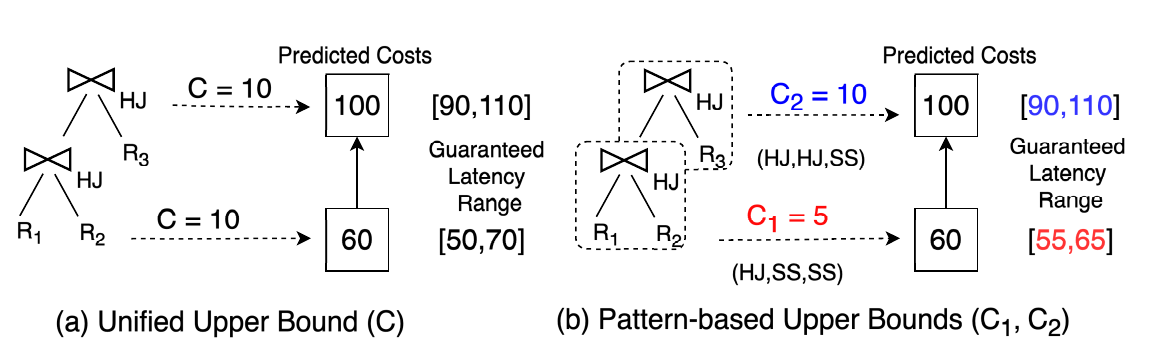}
    \vspace{-8pt}\caption{\revised{Bounded} Latency Ranges with Different Types of  Upper Bound ($C$).}
    \label{fig:upper_bound_modes}\vspace{-10pt}
\end{figure}

Figure~\ref{fig:upper_bound_modes} shows an example of using both unified and pattern-based upper bounds to calculate the \revised{bounded} latency ranges for one testing query plan. Here, we assume a white-box {\lqo} that constructs the plan from the bottom up. Initially, it constructs a Hash Join (HJ) at the first level, with Sequential Scan (SS) operations as left and right children. This parent-children pattern is labeled as (HJ, SS, SS)~\footnote{We assume that the roles of the left and right child operators are not interchangeable. Therefore, the order of children in any pattern is important, i.e., (HJ, SS, HJ) and (HJ, HJ, SS) are different patterns.}. Similarly, the partial plan in the second level has the (HJ, HJ, SS) pattern. In this example, the {\lqo} predicts 60 and 100 costs for these two partial plans. In the case of using unified upper bound (Figure~\ref{fig:upper_bound_modes}~(a)), we use a single value $C = 10$ is applied, resulting in latency ranges of \([50, 70]\) and \([90, 110]\) for the first and second partial plans, respectively. In the case of using pattern-based upper bounds (Figure~\ref{fig:upper_bound_modes}~(b)), two different values $C_1 = 5$ and $C_2 = 10$ are used, resulting in latency ranges of \([55, 65]\) and \([90, 110]\) for the (HJ, SS, SS) and (HJ, SS, HJ) patterns, respectively. 

%\textbf{Marginal Coverage Guarantees.} We would like to highlight that the guarantee in the equation is marginal, referring to the randomness across both the test and calibration data points \(R^{(0)}, R^{(1)},\\ \dots, R^{(K)}\), rather than conditional on the specific values of the calibration set \(R^{(1)}, \dots, R^{(K)}\). In other words, \(\text{Prob}(\cdot)\) reflects the randomness arising from the draw over all random variables. 

%Based on the previous discussion, we conclude the core formula we use in our framework as follows:
%\[
%P(\lVert t^{(i)} - \hat{c}^{(i)} \rVert \leq C) \geq 1 - \delta. \tag{5}\label{eq:5}
%\]
%where \(i\) means the data point index in our data series. We construct \(C\) based on the calibration set and then apply this \(C\) and \(\delta\) to other data points to provide the same guarantees. In our framework, this conformal prediction formula offers us guarantees with confidence.

\vspace{-8pt}
\section{{\cp}-based Runtime Verification for White-Box {\lqo}s Plan Construction}
\label{sec:cp_based_runtime_verification}

{\small
\begin{algorithm}[hbtp]
\caption{Constructing a List of Upper Bound(s) $\mathbf{C}$ on the Latency-Cost Non-Conformity Scores}
\label{alg:cp_upper_bound}
\begin{algorithmic}[1]
\Require  List of sorted latency-cost non-conformity scores~$R$, Uncertainty probability~$\delta \in [0, 1]$, Upper bound type~$T \in \{$\texttt{Unified}, \texttt{Pattern-based}$\}$    
\Ensure List of upper bound(s) $\mathbf{C}$
\State $K \gets$ length of sorted $R$

\If{$K < \frac{1 - \delta}{\delta}$} \Comment{Calibration scores are not enough}
    \State Get more $R$ scores from calibration
\Else 
    \If{$T$ is $\texttt{Unified}$} 
            \State $p \gets \lceil (K + 1)(1 - \delta) \rceil - 1$ \Comment{$(1-\delta)$th quantile index}
            \State $\mathcal{U} \gets R^{(p)}$ \Comment{$(1-\delta)$th quantile of $R$}
            \State $\mathbf{C} \gets \{\mathcal{U}\}$  \Comment{List has one unified upper bound}
    \Else \Comment{Upper bounds will be calculated based on patterns}
        \ForAll{parent-children patterns in calibration}
            \State $R_{patt} \gets$ List of sorted $R$ of the current pattern
            \State $n_{patt} \gets$ Size of $R_{patt}$

            \If{$n_{patt} < \frac{1 - \delta}{\delta}$} 
                \State Get more $R_{patt}$ scores from calibration
            \Else
                \State $p_{patt} \gets \lceil (n_{patt} + 1)(1 - \delta) \rceil - 1$ 
                \State $\mathcal{U}_{patt} \gets R^{(p_{patt})}$ 
                \State $\mathbf{C} \gets \mathbf{C} \cup \{\mathcal{U}_{patt}\}$ 
            \EndIf
        \EndFor
    \EndIf 
    \State\Return $\mathbf{C}$ 
\EndIf
\end{algorithmic}
\end{algorithm}
}

Earlier (Section~\ref{sec:cp_for_lqo}), we showed how {\cp} can provide a \revised{bounded} latency range for partial or complete query plans, helping assess the uncertainty of {\lqo} decisions before execution. Here, we aim to go further by exploring the use of {\cp} to early detect any performance constraint violations during the plan construction process of \textit{white-box} {\lqo}s (e.g.,~\cite{balsa, neo}), based solely on the constructed partial plans so far and before the full plan is completed. 

Suppose \(\mathcal{D}\) is an unknown distribution over the query plans generated by a white-box {\lqo}. Let \(X := (X_0, X_1, \dots) \sim \mathcal{D}\) represent a random query plan generated by the {\lqo}, where \(X_{\tau}\) is a random variable denoting the state of the generated partial plan at step~$\tau$ (e.g., predicted cost or actual latency). Then, we can formally define the white-box {\lqo} runtime verification problem as follows:\\

%Also, assume the availability of independent $\mathcal{Q}^{Tra}$ and $\mathcal{Q}^{Cal}$ query workloads for training and calibrating the {\lqo}, respectively.

\vspace{-8pt}

\textbf{Definition~1 (The White-Box {\lqo} Runtime Verification Problem).} \textit{Assuming a white-box {\lqo} (e.g.,~\cite{balsa}) and a testing query~$q$ that this {\lqo} already finished constructing its partial plans till step $\tau$ and is still running, we aim to \underline{verify} whether all generated partial plans by this {\lqo} (past and future) result in a complete plan, represented by $X$, that satisfies a user-defined STL-based performance constraint~$\phi$ with a confidence level $1-\delta$, i.e., $\text{Prob} (X \models \phi) \geq 1 - \delta$, where $\delta \in [0, 1]$ is a constraint violation probability.} \\

\vspace{-8pt}

Let \(x:= (x_0, x_1, \ldots)\) be the realization of \(X := (X_0, X_1, \dots)\), where \(x_{\text{obs}} := (x_0, \ldots, x_\tau)\) represents the constructed partial plans till step $\tau$ and \(x_{\text{un}} := (x_{\tau+1}, x_{\tau+2}, \ldots)\) represents the future unknown partial plans that will be predicted. Since existing \textit{white-box} {\lqo}s (e.g.,~\cite{balsa, neo}) predict one partial plan at a time, then we can estimate the realization $x$ at step $\tau$, with its constructed plans so far (i.e., $x_{\text{obs}}$) and next prediction at step $\tau+1$ as follows:
\[
\hat{x} := (x_{\text{obs}}, \hat{x}_{\tau+1 \mid \tau}) \label{3.2-1}\tag{7}
\]
As described in Definition~1, our goal is to verify the quality of the white-box {\lqo}'s complete query plan, represented by $X$, against a user-defined {\stl} specification \(\phi\). We can use robust semantics \(\rho^\phi(.)\) (check Section~\ref{sec:background_stl}) to achieve that. First, we define \(\rho^\phi(X)\) to indicate how robustly the specification \(\phi\) is satisfied with the complete query plan, and \(\rho^\phi(\hat{x})\) to denote the estimate of this robustness we obtained so far based on the observations \(x_{\text{obs}}\) and the prediction \(\hat{x}_{\tau+1 \mid \tau}\). Then, according to~\cite{cpstl23,prtstl23}, we can use {\cp} (Equation~\ref{2.2-1}) to define an upper bound \(C\) on the difference between the actual robustness \(\rho^\phi(X)\) of the complete query and the estimate of this robustness \(\rho^\phi(\hat{x})\) till step $\tau$ such that:  

\[
\text{Prob}(\rho^\phi(\hat{x}) - \rho^\phi(X) \leq C) \geq 1 - \delta \tag{8}\label{3.2-2}
\]

This upper bound \footnote{\revised{Note that upper bounds on the robustness values can be adjusted in the distribution shift cases using the same approach in Section~\ref{sec:cp_for_lqo_dynamic}.}} can be easily obtained from the calibration query workload $\mathcal{Q}^{Cal}$ by calculating the following non-conformity score $R^{(i)}$ for each partial plan in each calibration query $q_i \in \mathcal{Q}^{Cal}$:
\[
R^{(i)} := \rho^\phi(\hat{x}^{(i)}) - \rho^\phi(x^{(i)}) \label{3.2-3}\tag{9}
\]
where $x^{(i)}$ is the realization of $X$ for query $q_i$ (i.e., actual latencies and predicted costs for all partial plans in $q_i$) and \(\hat{x}^{(i)}\) is the estimate of this realization till step $\tau$ only (i.e., \(\hat{x}^{(i)} := (x_{\text{obs}}^{(i)}, \hat{x}_{\tau+1 \mid \tau}^{(i)})\)). Given that, we can define the following condition to verify whether the {\lqo} satisfies $\phi$ or not.\\

\vspace{-8pt}

\textbf{Lemma~2 (The White-Box {\lqo} Runtime Verification Condition).} \textit{Given a testing query $q$ that uses {\lqo} to generate its plan, represented by \(X\), and with $\hat{x} := (x_{\text{obs}}, \hat{x}_{\tau+1 \mid \tau})$ realizing the constructed and predicted partial plans at step $\tau$, an {\stl} constraint $\phi$, a robust semantics measure \(\rho^\phi(.)\) for this $\phi$ constraint, and a constraint violation probability $\delta \in [0, 1]$. Then, we can \underline{guarantee} that these constructed and predicted partial plans $\hat{x}$ so far will result in a complete plan that satisfies the constraint $\phi$ with a confidence level $1-\delta$, i.e., $\text{Prob} (X \models \phi) \geq 1 - \delta$, only if the robust semantics defined over these partial plans \underline{$\rho^\phi(\hat{x})$ is larger than $C$}, where $C$ is an upper bound calculated at a confidence level $1-\delta$ and using equations~\ref{3.2-2} and~\ref{3.2-3}. Otherwise, the resulting complete plan will cause a violation.}\\

\vspace{-8pt}

\textbf{Proof.} By reformulating Equation~\ref{3.2-2}, we can obtain:
\[
P(\rho^\phi(X) \geq \rho^\phi(\hat{x}) - C) \geq 1 - \delta\label{3.2-4}\tag{10}
\]
If \(\rho^\phi(\hat{x}) > C\), it implies:
\[
P(\rho^\phi(X) > 0) \geq 1 - \delta\label{3.2-5}\tag{11}
\]
which, according to Section~\ref{sec:background_stl}, further implies:
\[
P(X \models \phi) \geq 1 - \delta\label{3.2-6}\tag{12}
\]
because \(\rho^\phi(X) > 0\) directly implies that \(X \models \phi\). Note that changing the constraint specification \(\phi\) and/or the robust semantics measure \(\rho^\phi(.)\) does not require retraining the white-box {\lqo} to obtain valid guarantees because its prediction decisions, i.e., partial plans, are agnostic to any constraint specification.

\subsection{Framework Overview}
\begin{figure}
    \centering
    \includegraphics[width=\linewidth]{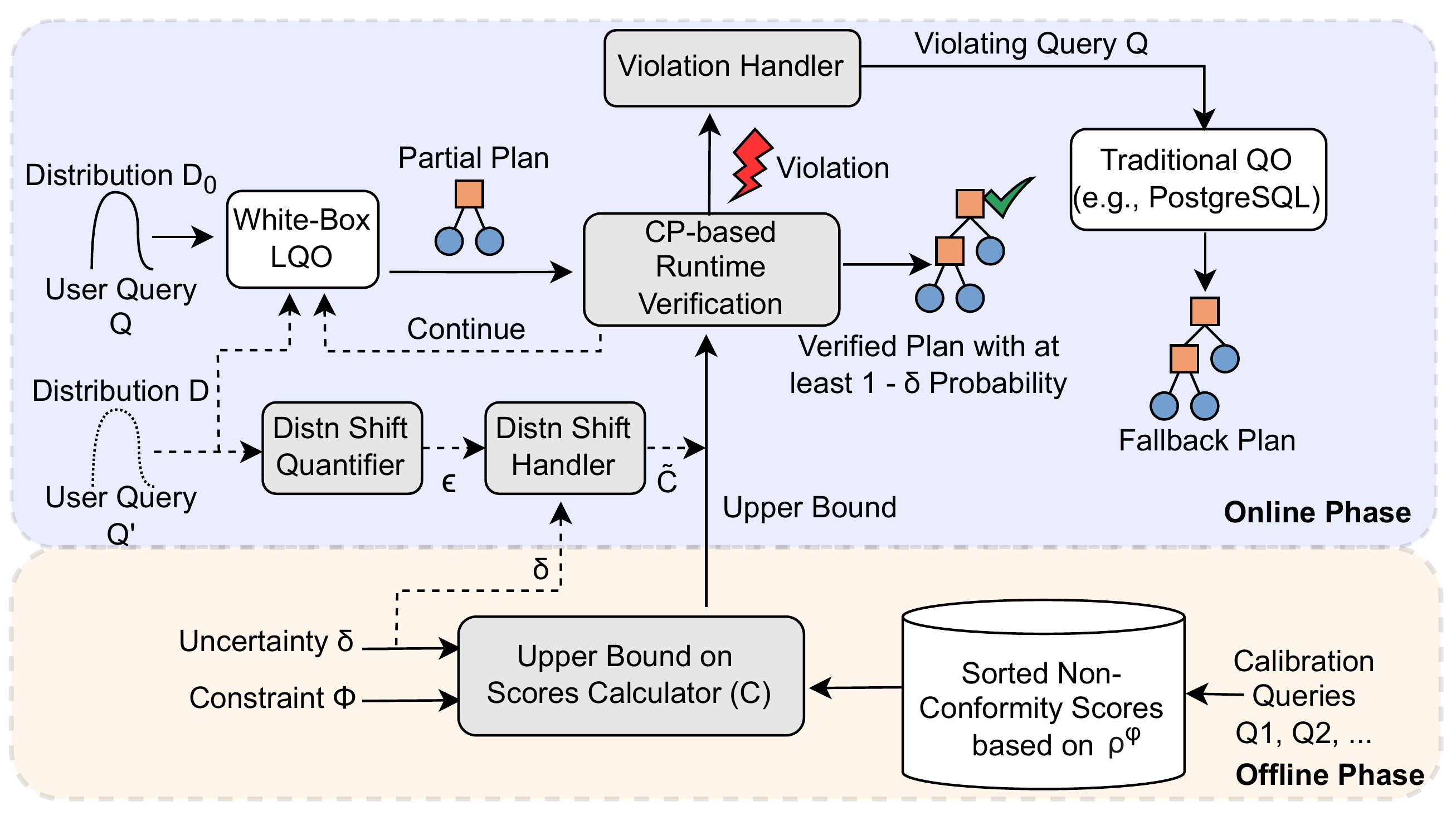}
    \vspace{-8pt}\caption{{\cp}-based Runtime Verification Framework.}
    \label{fig:cp4lqo_runtime_verification}\vspace{-18pt}
\end{figure}

Figure~\ref{fig:cp4lqo_runtime_verification} presents an overview of our {\cp}-based runtime verification framework, which detects violations of user-defined performance constraints $\phi$ in the plans being constructed by white-box {\lqo}s. 
  
\noindent\textbf{Offline Phase.} Similar to our bounded latency range framework (Section~\ref{sec:cp_latency_guarantee_framework}),  we start by constructing and sorting a set of non-conformity scores, obtained from the calibration queries and their partial plans. However, instead of constructing latency-cost-based scores (Equation~\ref{3.1-1}), we compute scores based on the difference between the actual robustness \(\rho^\phi(X)\) of queries and their estimated robustness \(\rho^\phi(\hat{x})\) at each partial plan step, assessing compliance with constraint $\phi$ using robustness measure $\rho^\phi(.)$ (Equation~\ref{3.2-3}). \revised{These scores are then sorted and used to compute any upper bound, whether $C$ in the static case or $\tilde{C}$ in the distribution shift case, at a user-defined confidence level $1-\delta$ (Equation~\ref{3.2-2}) as discussed previously in Section~\ref{sec:cp_latency_guarantee_framework}}.

 \noindent\textbf{Online Phase.} When a user submits a testing query, the white-box {\lqo} starts to incrementally build the plan, adding one partial plan at a time. At each step $\tau$, the runtime verification module uses the upper bound $C$ and the estimated robustness \(\rho^\phi(\hat{x})\) (representing all partial plans constructed up to $\tau$ and the expected one at $\tau + 1$) to check if \(\rho^\phi(\hat{x}) > C\) (Lemma~2). If this condition holds, the {\lqo} proceeds to construct the next partial plan at step $\tau + 1$. Otherwise, a violation is detected (e.g., exceeding a latency threshold). As a result, the violation handler discards the current plan under construction and sends the query to be re-planned by a traditional query optimizer (e.g., PostgreSQL~\cite{postgres2024}). This has been shown to be an effective solution, as highlighted in earlier works (e.g.,~\cite{bao}) and confirmed by our experimental evaluation (Section~\ref{sec:evaluation}). The intuition here is that re-planning the query with a traditional optimizer and running it with the resulting average-performance plan incurs less overhead than executing a worst-case {\lqo}-generated plan. \revised{Note that in case there is a distribution shift in the testing queries from \(\mathcal{D}_0\) to \(\mathcal{D}\), we construct the adjusted upper bound \(\tilde{C}\) as in the online phase of our bounded latency range framework (Section~\ref{sec:cp_latency_guarantee_framework}).}
\vspace{-8pt}\section{{\cp}-Guided Plan Search in White-Box {\lqo}s}
\label{sec:cp_guided}

In this section, we provide a simple yet effective approach for using {\cp} to steer the decision-making process in white-box {\lqo}s. Unlike sections~\ref{sec:cp_for_lqo} and~\ref{sec:cp_based_runtime_verification},  which focused on using {\cp} to obtain \revised{bounded} latency ranges for generated plans or to detect violations during the plan construction process (triggering a fallback to traditional optimizers), this section presents a {\cp}-guided plan search algorithm designed to improve the quality of generated plans rather than just verifying them. %\sout{Specifically, we provide a variant of the beam search algorithm\mbox{~\cite{lowerre1976harpy, balsa}}}. 
\revised{Specifically, this algorithm utilizes {\cp}-derived upper bounds on the actual latency of partial plans (Equation~\ref{4.3-2}), to heuristically guide the plan search space navigation}.

%\subsection{Algorithm Overview}
\noindent\textbf{Intuition.} White-box {\lqo}s, such as {\balsa}~\cite{balsa} and {\neo}~\cite{neo}, use learned cost predictors to search over the space of partial plans at each step, aiming to identify the plan with the lowest predicted cost. However, since the space of all partial plans at any step is far too large to exhaustively search, these {\lqo}s typically find this plan heuristically by sorting predicted costs and then selecting the plan with the lowest cost. However, relying on the predicted costs can lead to sub-optimal plans if these predicted costs do not closely align with the actual latencies. To address this, we propose leveraging the {\cp}-\revised{bounded} upper bounds on actual latency, which was discussed in Section~\ref{sec:cp_for_lqo}, to guide the search for optimal partial plans at each step.

\noindent\textbf{{\cp}-Guided Plan Search Algorithm.} Recall that for any partial plan at step $\tau$, we can compute an upper bound $U_{\tau}$ on its actual latency $t_{\tau}$ as $\hat{c}_{\tau} + C$ (right inequality in Equation~\ref{4.3-2}), where $\hat{c}_{\tau}$ represents the predict cost of this partial plan and $C$ is the upper bound on the error between $t_{\tau}$ and $\hat{c}_{\tau}$, calculated at a user-defined confidence level of $1 - \delta$. \revised{Based on this latency upper bound~$U_{\tau}$, we propose a generic {\cp}-guided plan search algorithm that is compatible with basic plan search (BPS) algorithms. Algorithm~\ref{alg:cp-guided} shows the details.} We first initialize a priority queue with a set of partial plans, each representing a scan operation over a relation in the user query. We also initialize \texttt{complete\_plans} to store the complete plans as they are identified (lines \textit{1-2}). \revised{At each iteration of the while loop, a partial plan, referred to as \texttt{state}, is retrieved from the \revised{priority queue} according to BPS's logic for selecting the next plan (lines \textit{3-4}). This selection logic may involve fetching the partial plan with the minimum cost (Best-First Search), iterating over each state in the current queue (Beam Search~\cite{lowerre1976harpy}), or using other strategies. If \texttt{state} forms a complete plan, it is added to the set of complete plans (lines \textit{5-8}).} Otherwise, the search continues from the current partial plan, \texttt{state}, by calling \texttt{Explore(.)}, which generates a new set of partial plans along with their predicted costs

%\sout{Beam Search\mbox{~\cite{lowerre1976harpy}}, a commonly used heuristic plan search algorithm in white-box {\lqo}s (e.g.,\mbox{~\cite{balsa,rejoin18}}). }
% Algorithm~\ref{alg:cp-guided} shows the details of our proposed variation. We first initialize 
%\revised{\sout{the \texttt{beam}, which is essentially}} 
% a priority queue, with a set of partial plans, each representing a scan operation over a relation in the user query. We also initialize  \texttt{plans} to store the complete plans as they are identified (lines \textit{1-2}). At each iteration in the while loop, the best partial plan, referred to as \texttt{state}, is retrieved from the 
%\sout{beam} 
% \revised{priority queue} (lines \textit{3-4}). If \texttt{state} forms a complete plan, it is added to the set of complete plans (lines \textit{5-8}). Otherwise, the search proceeds from the current partial plan, \texttt{state}, by calling \texttt{Explore(.)}, which retrieves a new set of partial plans with their predicted costs.

For each new partial plan (\texttt{stateNew}), we use Algorithm~\ref{alg:cp_upperbound} (described later) to compute its latency upper bound $U_{\text{stateNew}}$ based on its predicted cost $\hat{c}_{\text{stateNew}}$ and the corresponding pattern-based upper bound on the latency-cost scores from $\mathbf{C}$. \revised{Then, these new states, \texttt{stateNew}, along with their corresponding values \(\hat{c}_{\text{stateNew}}\) and \(U_{\text{stateNew}}\), follow BPS's logic for inserting new plans (lines \textit{10-12}). This insertion logic may involve directly adding the new plans to the priority queue (non-optimized plan search). Alternatively, it may involve shrinking the queue to a specific size after inserting multiple plans, retaining only the smallest \(bLen\) plans for further exploration (Beam Search).
} The algorithm continues until $n$ complete plans are identified. Finally, these complete plans are sorted based on their latency upper bounds, and the top-ranked plan is selected as the \texttt{final} plan.

{\small
\begin{algorithm}[htbp]
\caption{\revised{{\cp}-Guided Plan Search}}
\label{alg:cp-guided}
\begin{algorithmic}[1]
\Require Learned cost predictor \texttt{LCP}, Pattern-based upper bounds $\mathbf{C} = \{C_{\text{1}}, C_{\text{2}}, \ldots\}$ from Algorithm~\ref{alg:cp_upper_bound}, Number of candidate complete  plans \texttt{n}, Basic plan search algorithm \texttt{BPS}.
\Ensure Top-ranked plan \texttt{final}
\State \revised{\texttt{queue} $\gets$ Partial plans initialized with scans over relations}
\State \texttt{complete\_plans} $\gets$ [] 
\While{\texttt{len(complete\_plans)} $<$ \texttt{n} \textbf{and} \revised{\texttt{queue} is not empty}}
    % \State \texttt{state} $\gets$ \texttt{queue.pop()} 
    \State \revised{(\texttt{state}, $\hat{c}_{\text{state}}$) $\gets$ BPS.select\_next\_plan(queue)}
    \If{\texttt{state} is a complete plan}
        \State \texttt{complete\_plans.add(state)}
        \State \textbf{continue}
    \EndIf
    \State List of (\texttt{stateNew}, $\hat{c}_{\text{stateNew}}$) $\gets$ \texttt{Explore(\texttt{LCP}, state)}
    \ForAll{pair in List of (\texttt{stateNew},  $\hat{c}_{\text{stateNew}}$)}  
        \State \small{$U_{\text{stateNew}}$ $\gets$ LatencyUpperBound (\texttt{stateNew}, $\hat{c}_{\text{stateNew}}$, $\mathbf{C}$)}
        % \State Add pair (\texttt{stateNew}, $U_{\text{stateNew}}$) to \texttt{queue} 
        \State \revised{BPS.insert\_plan(queue, \texttt{stateNew}, $\hat{c}_{\text{stateNew}}$, $U_{\text{stateNew}}$)} 
    \EndFor
\EndWhile
\State Sort \texttt{complete\_plans} by $U_{\text{state}}$ values in an ascending order
\State \texttt{final} $\gets$ \texttt{complete\_plans[0]}\\
\Return \texttt{final}
\end{algorithmic}
\end{algorithm}
}
\noindent\textbf{Latency Upper Bound Calculation.} Algorithm~\ref{alg:cp_upperbound} shows how the latency upper bound is calculated. It first extracts the parent-children pattern of the input partial plan. Then, it retrieves the upper bound on the latency-cost non-conformity scores corresponding to this pattern, referred to as \texttt{latencyCostUpperBound}, from $\mathbf{C}$. In case not found, \texttt{latencyCostUpperBound} is assigned the maximum value in $\mathbf{C}$. This is important to guarantee the plan selection quality during the beam search in Algorithm~\ref{alg:cp-guided} because when a pattern is not found, the value of its latency upper bound $U_{\tau}$ becomes very large, due to the addition of $\max(\mathbf{C})$ to the predicted \texttt{cost}, and hence its priority to be selected during the \revised{{\cp}-guided} search compared to other partial plans with patterns having values in $\mathbf{C}$ (i.e., trusted partial plans) is very low.  %\toprof{is low but still maintains possibility.}

{\small
\begin{algorithm}[hbtp]
\caption{Calculate {\cp} \revised{Bounded} Latency Upper Bound}

\label{alg:cp_upperbound}
\begin{algorithmic}[1]
%\Require Pattern hashmap \texttt{cp\_hashmap} = \{\text{pat}: $C_{\text{pat}}$\}
\Function{LatencyUpperBound}{\texttt{state}, \texttt{cost}, $\mathbf{C}$}
\State \texttt{pattern} $\gets$ \texttt{ExtractPattern(state)} 
\If{\texttt{pattern} has an upper bound $C_{\text{pat}} \in \mathbf{C}$}
    \State \texttt{latencyCostUpperbound}$\gets$ $C_{\text{pat}}$
\Else \Comment{\texttt{pattern} is not seen before}
    \State \texttt{latencyCostUpperbound} $\gets \max(\mathbf{C}) $
\EndIf
\State \Return \texttt{cost} + \texttt{latencyCostUpperbound}
\EndFunction
\end{algorithmic}
\end{algorithm}
}
\vspace{-9pt}\section{Experimental Evaluation}
\label{sec:evaluation}

We evaluated our {\cp}-based frameworks using different benchmarks and multiple prototypes to address the following questions: (1) How effective are the multi-granularity {\cp}-based latency guarantees (Section~\ref{sec:eval-multi-granularity})? \revised{(2) How effective does our adaptive {\cp} handle distribution shift? (Section~\ref{sec:evaluatioin-adaptive-cp})?} (3) How effective is our runtime verification (Section~\ref{sec:eval-runtime-verification}) (4) How much performance gain can be achieved through effective violation detection and handling (Section~\ref{sec:eval-violation-detection-and-handling})? (5) What benefits does {\cp}-guided plan search provide in terms of plan quality and planning time (Section~\ref{sec:eval-cp-guided})? \revised{(6) What is the sensitivity of the hyper-parameters of our {\cp}-based approach and their effects on the {\lqo} verification process (Section~\ref{sec:eval-hyper-parameter}) }

% \tohanwen{You should set the performance questions you want to answer and refer to the sections answering them}
\vspace{-6pt}\subsection{Experimental Setup}
% For the JOB workload, Balsa was trained using 69 queries over 50 iterations and 100 epochs, with an additional 50 queries used for calibration and testing. For the TPCH workload, Balsa was trained on 60 queries, also over 50 iterations and 100 epochs, with 70 queries used for calibration and testing.
\subsubsection*{\textbf{{\cp} Integration with three {\lqo}s}}
\noindent\textit{\underline{Balsa~\cite{balsa}}} integrates a learned cost predictor and beam search, storing ranked potential sub-plans to construct a complete plan. \revised{We choose {\balsa} as our default \textit{white-box} {\lqo} due to its superior performance over other {\lqo}s in this category (e.g., NEO~\cite{neo}) as shown in many studies~\cite{base23,balsa,zhu2023lero}}. 
%\textcolor{red}{\sout{Given Balsa’s successful latency fine-tuning and its \textit{white-box} {\lqo} features, we use it as the default {\lqo} in all experiments, unless otherwise mentioned.}}
%\revised{Balsa's \textit{white-box} {\lqo} characteristics make it well-suited for all our proposed methods.} 
For latency bounds experiments, we verify both the unified-based and pattern-based upper bounds. To perform runtime verification, we calculate the robustness \(\rho^\phi(\hat{x})\) (see Section~\ref{sec:cp_based_runtime_verification}) and compare it with the corresponding upper bound. If a violation is detected, our violation handler addresses it by reverting to PostgreSQL~\cite{postgres2024}. Finally, we perform {\cp}-guided plan search to compare with the original Balsa models trained with different iterations.

\noindent\textit{\underline{Lero~\cite{zhu2023lero}}} generates multiple candidate query plans and uses a learned oracle to rank them. The oracle applies pairwise comparisons to predict the more efficient plan, selecting the top-ranked one as the final output. Since Lero operates as a \textit{black-box} {\lqo} without directly accessible cost information, we use PostgreSQL's predicted costs as a reference and the actual latency to construct {\cp} model. At this stage, the predicted cost \(\hat{c}\) is available, and we use {\cp} to derive a guaranteed range for the actual runtime \(t\) based on \(\hat{c}\). We then apply {\cp} models at various granularities to estimate the guaranteed range for the entire plan, individual levels, and identified patterns.

% RTOS represents a distinct branch of learned query optimization focused on join order selection. RTOS leverages a Reinforcement Learning framework in conjunction with Tree-LSTM~\cite{tai2015improved} to effectively capture the structural information of query plans. Unlike earlier learned optimizers that primarily encode join trees using fixed-length vectors, RTOS employs a Tree-structured Long Short-Term Memory (Tree-LSTM) model, which is capable of adapting to changes in the database schema.
\noindent\textit{\underline{RTOS~\cite{yu2020reinforcement}}} focuses on join order selection, leveraging a DRL framework in conjunction with Tree-LSTM~\cite{tai2015improved} to effectively capture the structural information of query plans. RTOS outputs a join ordering hint, which we then inject into PostgreSQL to generate a complete query plan. As another representative of the \textit{black-box} {\lqo}, RTOS is similar to Lero’s {\cp} integration in that we use PostgreSQL’s predicted cost as a reference. Given that RTOS has less control over the selection of plan operators (such as Sequential Scan), we primarily use RTOS to validate our {\cp}-based latency guarantee framework. 

\revised{Notably, existing {\lqo}s face significant limitations in handling distribution shifts, restricting their ability to generalize across different workloads. They are either hard-coded to specific schemas in their open-source implementations or require processing all training queries upfront to define the model structure, making it impossible to optimize unseen queries dynamically.  For instance, the open-source version of RTOS~\cite{yu2020reinforcement} is hard-coded to IMDB table schemas, restricting its use to the JOB~\cite{qoeval15} and JOBLight-Train~\cite{Kipf2019GroupBy} workload only. Therefore, to explicitly enable {\lqo}s to operate across distributions, we modified their prototype frameworks to support changing distributions.
}

% is another representative of the \textit{black-box} model. RTOS outputs a join ordering hint, which we then inject into PostgreSQL to generate a complete query plan. Similar to Lero's {\cp} prototype, we use PostgreSQL's predicted cost as a reference, and apply the function $f(\hat{c}) = \hat{c} / 50$ for normalization. Given that RTOS has less control over the selection of plan operators (such as Sequential Scan), we primarily use RTOS to validate our {\cp}-based latency guarantee framework.

% Through its machine learning model, RTOS generates a join ordering, serving as a hint for the query optimizer to construct the entire query plan. During the actual query plan construction, we can apply conformal prediction to evaluate the quality of the predicted join ordering. The predicted cost \(\hat{c}\) reflects RTOS's join order selection effectiveness. Based on \(\hat{c}\), we then proceed to perform both guarantees and runtime verification, ensuring the robustness and efficiency of the query execution. We use function $f(\hat{c})=\hat{c}/50$ as the normalization.
\vspace{-6pt}\subsubsection*{\textbf{Benchmarks}}

\revised{We evaluate the integration of {\cp} with these {\lqo}s on four widely used benchmarks - Join Order Benchmark (JOB)~\cite{qoeval15}, Cardinality Estimation Benchmark (CEB)~\cite{Negi2021FlowLoss}, JOBLight-train~\cite{Kipf2019GroupBy}, and TPC-H~\cite{tpch}. For the static case evaluation, we use JOB and TPC-H workloads.} JOB workload consists of 113 analytical queries over a real-world dataset from the Internet Movie Database. These queries involve complex joins and predicates, ranging from 3-16 joins, averaging 8 joins per query. For our experiments, we select 33 queries for model training, while the remaining 80 queries are used for calibration and testing. TPC-H features synthetically generated data under a uniform distribution. We use a scale factor of 1 and templates for queries 3, 5, 7, 8, 10, 12, 13, and 14 to generate workloads, creating 130 queries with varying predicates. Of the generated queries, 60 are used for model training, while the remaining 70 are designated for calibration and testing. \revised{For the distribution shift case evaluation, we use JOBLight-train and CEB workloads along with JOB. JOBLight-train consists of synthetically generated queries with 3-table joins. CEB employs hand-crafted templates and query generation rules to construct challenging large queries.}

\vspace{-6pt}\subsubsection*{\textbf{Hardware and Settings}}
The experiments related to Balsa and RTOS were conducted on an Ubuntu 22 machine with an 8-core Intel Xeon D-1548 CPU @ 2.0GHz and 64 GB of RAM. The experiments related to Lero were conducted on an Ubuntu 22 machine with a 10-core Intel Xeon Silver 4114 @ 2.2GHz and 64 GB of RAM.

\vspace{-6pt}\subsubsection*{\textbf{{\cp} Empirical Coverage (EC)}}
% To empirically validate marginal guarantees of the form \ref{2.2-1}, we perform the following experiment \(N\) times: we sample one test query \(R^{(0)}\) and \(K\) calibration trajectories \(R^{(1)}, \dots, R^{(K)}\) from the distribution \(\mathcal{R}\). For each experiment, we check whether the condition \(R^{(0)} \leq C(R^{(1)}, \dots, R^{(K)})\) is satisfied. 

% We then compute the empirical coverage, defined as the ratio of the number of times the condition \(R^{(0)} \leq C(R^{(1)}, \dots, R^{(K)})\) holds, divided by the total number of experiments \(N\). As \(N \to \infty\), we observe that the empirical coverage converges to \(1 - \delta\).

% Formally, for \(R_n^{(0)}, R_n^{(1)}, \dots, R_n^{(K)} \sim \mathcal{R}\), with \(n \in \{1, \dots, N\}\), the empirical coverage (EC) is computed as:

% \[
% EC := \frac{1}{N} \sum_{n=1}^{N} \mathbbm{1}(R_n^{(0)} \leq C(R_n^{(1)}, \dots, R_n^{(K)}))
% \]
% where \(\mathbbm{1}(R_n^{(0)} \leq C(R_n^{(1)}, \dots, R_n^{(K)}))\) is the indicator function that takes the value 1 if the condition \(R_n^{(0)} \leq C(R_n^{(1)}, \dots, R_n^{(K)})\) holds, and 0 otherwise.\toshashank{Change to general expression not only R0?, Ans: No, since R0 is a sample }

% Hanwen will update to calibration empirical coverage

To empirically validate the {\cp} marginal guarantees of Formula~\ref{2.2-1}, we conduct the experiment over $M$ iterations. For each iteration, we sample \(K\) calibration queries, $\{Q^{(1)}, \dots, Q^{(K)}\}$, and $N$ test queries, $\{Q_{1}^{(0)}, \dots, Q_{N}^{(0)}\}$. Then, we calculate $EC_{m}$ for iteration $m$ using the following formula:
\[
EC_{m} := \frac{1}{N} \sum_{n=1}^{N} \mathbbm{1}(R_{m,n}^{(0)} \leq C(R_m^{(1)}, \dots, R_m^{(K)})).\label{equ:ec}\tag{13}
\]
\vspace{-4pt}\noindent\textbf{Evaluation Metrics.} We focus on the following metrics: (1) \textit{Coverage:} $EC_{m}$, calculated as defined in Equation~\ref{equ:ec} for each sampling iteration $m$ and presented as a percentage. This value measures the validity condition on the test set when applying our constructed $C$, indicating how many test cases are successfully covered. (2)~\textit{Frequency Density:} Across all sampling iterations, we calculate the frequency of each coverage level. To more effectively display the data, we use Kernel Density Estimation (KDE) for density representation. \revised{Intuitively, a higher frequency density for a specific coverage indicates a greater likelihood of its occurrence during sampling.} (3) \textit{{\cp} Upper Bound $C$:} This is the {\cp} upper bound for latency-cost non-conformity scores (see Section~\ref{sec:cp_for_lqo}), used to compare the spans of non-conformity scores across different hyper-parameter settings. \revised{(4) \textit{Non-conformity Scores:} We display the distribution of non-conformity scores in the runtime verification context to visually validate how runtime constraints are satisfied.} (5) \textit{Execution Latency:} The actual execution latency, measured in milliseconds (ms), is used to assess the quality of generated query plans. (6) \textit{Planning Time:} The time taken to generate a query plan is used to evaluate the algorithm’s search efficiency during the planning.

\noindent\revised{\textbf{Default Parameters.}} \revised{Unless otherwise mentioned, in any experiment, we run multiple sampling iterations ($N = 1000$) to observe the empirical coverage. In each iteration, we randomly select a fixed-size calibration set to generate non-conformity scores and then construct $C$ based on the given $1 - \delta$. We set $\delta = 0.1$ and the calibration-test split to be 50\%-50\%. When validating a testing query, we perform the evaluation on each operator $i$ in the query for Unified-based upper bounds or each pattern $i$ (parent-children structure) for Pattern-based upper bounds.  Each instance is treated as a test step $i$, where we have the predicted cost $\hat{c}_{i}$ and the actual latency $t_{i}$. Combining $C$ with $\hat{c}_i$ to calculate the bounded range for actual latency $[\hat{c}_{i}-C, \hat{c}_{i}+C]$, then verify $t_{i} \in [\hat{c}_{i}-C, \hat{c}_{i}+C]$. For the complete test set, we calculate the coverage for this bounded range method, which ranges between 0 and 1. We normalize the predicted costs in Lero and RTOS cases with $f(\hat{c}) = \hat{c}/40$ and $\hat{c}/100$, respectively, to align these costs with actual latencies\footnote{\revised{Note that we do not normalize the predicted costs for {\balsa} as it aims to predict the expected latency of the generated plan. So costs and latencies are well-aligned.}}. }

%\revised{Based on the hyper-parameter selection section (check Section~\ref{sec:eval-hyper-parameter})}, we confirm the hyper-parameters as follows: . .

% We perform experiments covering the three levels of granularity for {\cp} mentioned in Section~\ref{sec:eval-multi-granularity-cp}. In this section, we use \(N=100\) iterations to calculate the empirical coverage. We will perform a random sampling process for each iteration's calibration set and test set. We put an uncertainty probability of \(\delta = 0.1\) to construct the {\cp} model. The non-conformity scores were calculated using Eq.~\(\ref{3.1-1}\).

\vspace{-6pt}\subsection{Bounded Range of Plan Actual Latency}\label{sec:eval-multi-granularity}

% \subsubsection{Whole-Plan {\cp}}

% \insights{All the three LQOs conform similarly. We need to note that the size of the data for whole plan is significantly low (For unified if each query contains 10 subplans we will have 80*10 datapoints for calibration + test where as for whole plan we will only have 40). This could be a reason for similar performance? Also an important point to note is ROTS does not change much but Lero and Balsa's Mean has moved toward the higher side, this can also be because of the data set size.}

% \begin{figure}[h]
%     \centering
%     \includegraphics[width=1\linewidth]{figures//1000iter//BalsaXImdb/wholePLan.png}
%     \caption{Balsa X JOB Whole Plan}
%     \label{fig:WholePlan}
% \end{figure}

% \begin{figure}[h]
%     \centering
%     \includegraphics[width=1\linewidth]{figures//1000iter//BalsaXImdb/LEROxIMDBWholePlan.png}
%     \caption{Lero X JOB Whole Plan}
%     \label{fig:enter-label}
% \end{figure}

% \begin{figure}[h]
%     \centering
%     \includegraphics[width=1\linewidth]{figures//1000iter//BalsaXImdb/RTOSxIMDBWholePlan.png}
%     \caption{RTOS x JOB Whole Plan}
%     \label{fig:enter-label}
% \end{figure}

% \subsubsection{Unified (Sub-Plan) {\cp}} 

\begin{figure}[h]
    \centering
    \begin{subfigure}{0.48\linewidth}
        \centering
        \includegraphics[width=\linewidth]{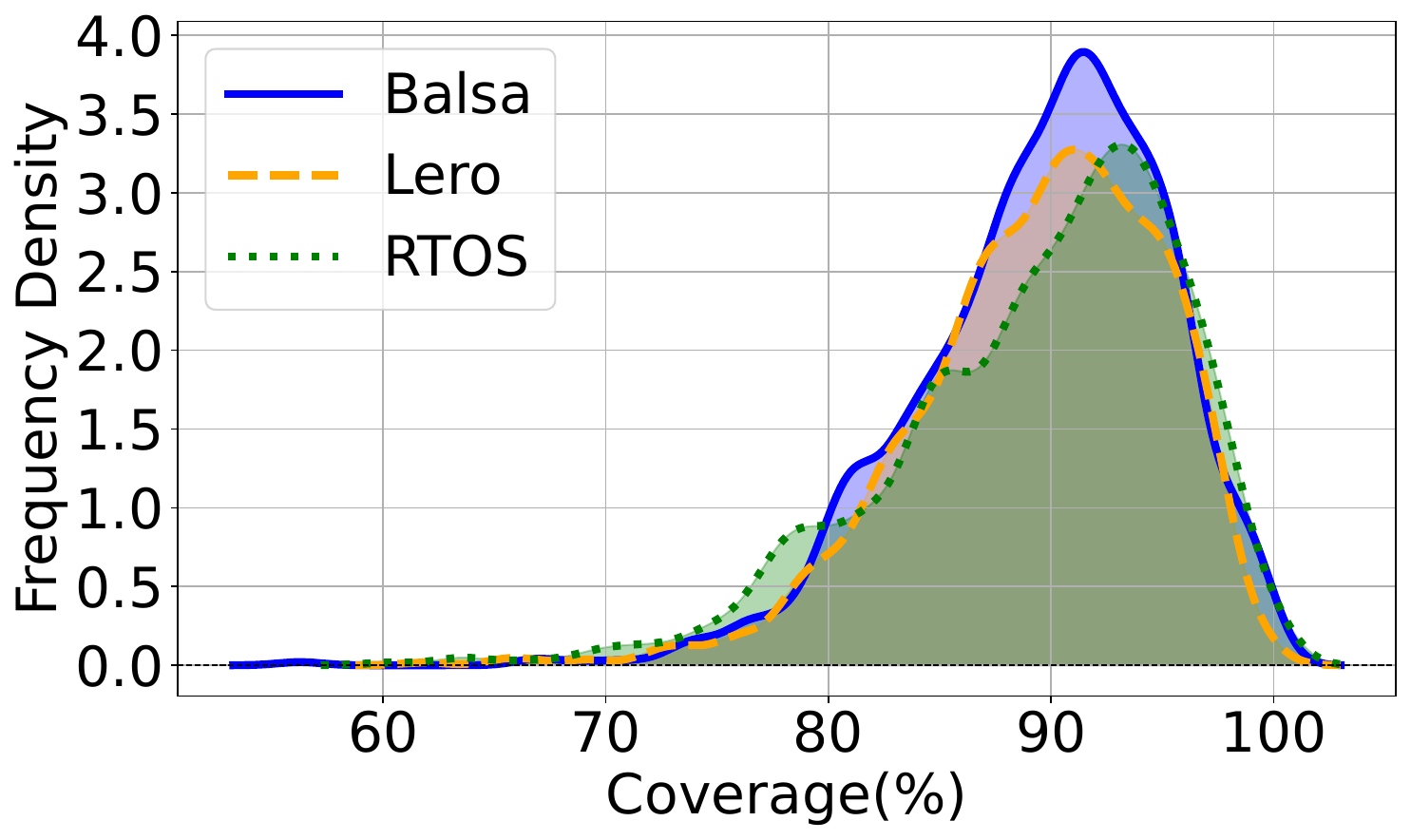}
        \vspace{-12pt}\caption{JOB}
        \label{fig:UnifiedCPCompareJOB}
    \end{subfigure}
    \hfill
    \begin{subfigure}{0.48\linewidth}
        \centering
        \includegraphics[width=\linewidth]{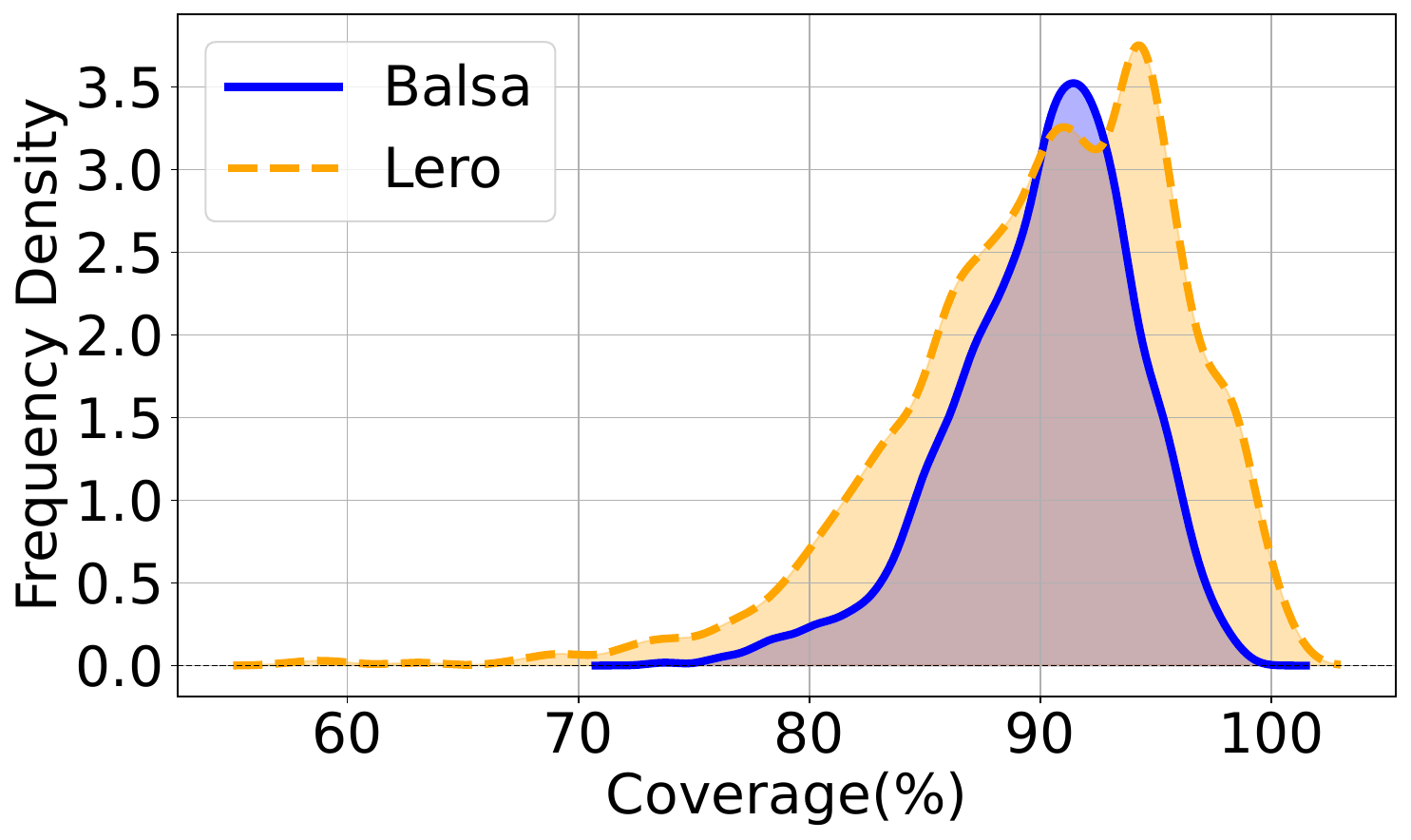}
        \vspace{-12pt}\caption{TPC-H}
        \label{fig:UnifiedCPCompareTPCH}
    \end{subfigure}
    \vspace{-8pt}\caption{Unified-cased Upper Bounds.}
    \label{fig:unified-cp}
\vspace{-10pt}
\end{figure}

\noindent\textbf{Unified-based Upper Bound.} We treat all the partial plans of a query plan equally with a single upper bound value for $C$ (check Section~\ref{sec:cp_latency_guarantee_framework}). Figure~\ref{fig:unified-cp} shows the empirical coverage in this case. We perform experiments on both the {\job} and {\tpch} workloads. \revised{According to Equation~\ref{4.3-2}, the {\cp} theory predicts that the most frequent coverage should be greater than $1 - \delta = 0.9$, as reflected by the peak of the curve in both graphs. For both workloads, the peak of all the curves demonstrates this trend, empirically validating the correctness of applying {\cp} with {\lqo}s}. In the JOB workload, we observe that Balsa and Lero show coverage\% more concentrated around 90\%, whereas RTOS exhibits a slightly relaxed coverage curve with a higher coverage peak in the middle. This variance could arise from differences in {\lqo}s' architecture. RTOS lacks partial plan level specific training, while Balsa and Lero perform more granular analysis on plans during training. Consequently, the non-conformity scores for operators in RTOS span a broader range than in Balsa and Lero. Even though $C$ derived from a sparse calibration space can adequately cover a dense test space, the reverse is less effective. This mismatch results in a higher coverage peak for RTOS but with relatively lower density.

\begin{figure}[h]
    \centering\vspace{-10pt}
    \begin{subfigure}{0.48\linewidth}
        \centering
        \includegraphics[width=\linewidth]{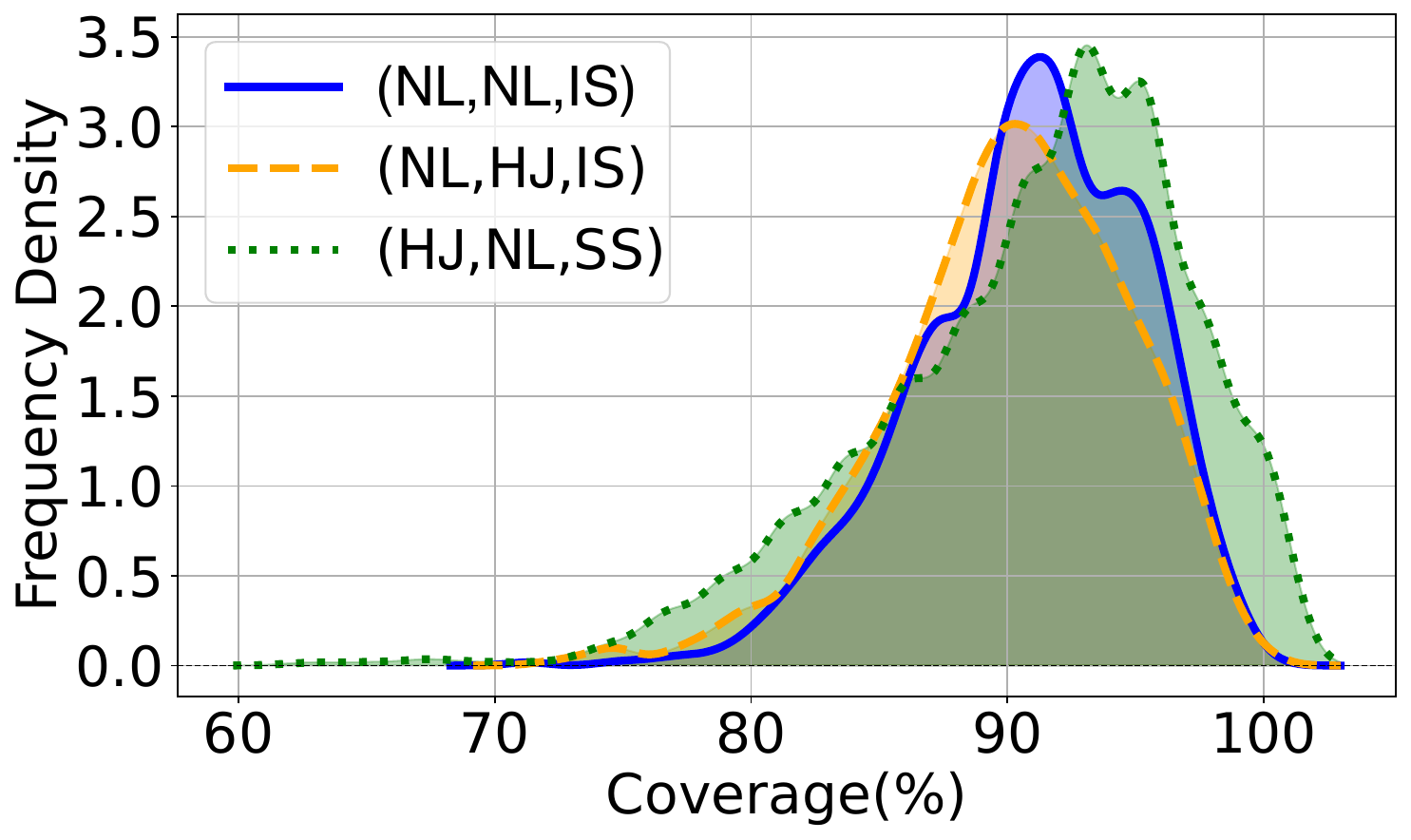}
        \vspace{-8pt}\caption{Top 3 Popular}
        \label{fig:Top3Patterns}
    \end{subfigure}
    \hfill
    \begin{subfigure}{0.48\linewidth}
        \centering
        \includegraphics[width=\linewidth]{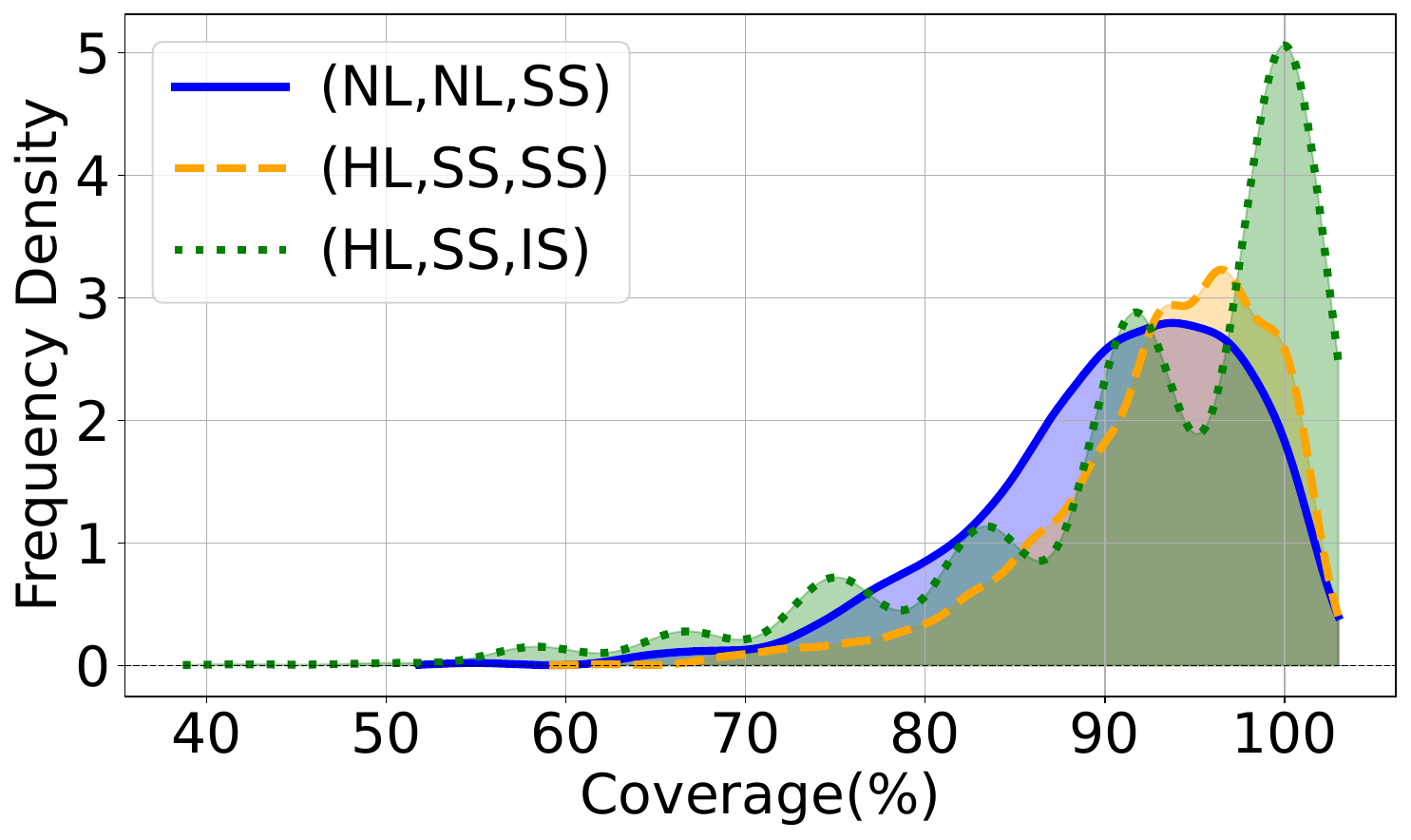}
        \vspace{-8pt}\caption{Least 3 Popular}
        \label{fig:Least3Patterns}
    \end{subfigure}
\vspace{-10pt}\caption{Valid Pattern-based Upper Bounds (Balsa on JOB)}
\label{fig:pattern-cp}
\vspace{-10pt}\end{figure}

\noindent\textbf{Pattern-based Upper Bound.}
Pattern-based upper bound provides finer granularity for generating a \revised{bounded} range. In this experiment, we examine the top 3 and least 3 frequently occurring patterns in Balsa on the JOB workload. Figure~\ref{fig:pattern-cp}~(a) displays the top 3 popular patterns: (NL, NL, IS), (NL, HJ, IS), and (HJ, NL, SS). The peak coverage reaches 0.9, with a mean \( C \) value of 3056 ms, indicating that Balsa’s actual latencies vary within a range of \(\pm3056\) ms. Figure~\ref{fig:pattern-cp}~(b) shows the least 3 popular patterns. Given that they have fewer appearances, which slightly exceeds the $K^{*}$ threshold, the curve is not as symmetric as the previous one. However, we also observe that the empirical coverage peak surpasses 90\%, indicating reliable, guaranteed latency. \revised{This also shows that the $\cp$ theory holds its ground when the value of $K$ is low yet greater than the $K^{*}$ threshold}.

% Fig.~\ref{fig:Top3Patterns} illustrates the results of Sub-structure {\cp} at a more detailed level. Instead of evaluating each operation in isolation without contextual awareness, we identified the three most frequently utilized patterns within Balsa’s optimization process for the JOB workload on the JOB dataset. One predominant pattern identified is the sequence (Hash Join, Hash Join, Seq Scan). We collected the costs and latencies of these patterns to construct non-conformity scores using the calibration queries. The latency guarantee was dynamically verified in real-time during query plan generation for the test queries. As shown in Fig.~\ref{fig:Top3Patterns}, 
\vspace{-6pt}\revised{\subsection{Adaptive {\cp} under Distribution Shift}\label{sec:evaluatioin-adaptive-cp}

We perform evaluations on Balsa~\cite{balsa} and RTOS~\cite{yu2020reinforcement} for distribution shift analysis. Our approach is inspired from existing works on distribution shift ~\cite{negi2023robust,wu2023addingdomainknowledgequerydriven,10.1145/3639293}, where the {\lqo}s are trained on one distribution and tested on another. Regarding the selection of distributions, we follow ~\cite{negi2023robust} and use the following distributions: JOB~\cite{qoeval15}, CEB~\cite{Negi2021FlowLoss}, JOBLight-train~\cite{Kipf2019GroupBy}.

% \subsubsection{Balsa}

%%%%%%%%%%
% We train Balsa and collect the calibration plans on JOB. We use a similar query split for Balsa training as in previous experiments. Specifically, we use 33 queries to train Balsa for 500 iterations. We run Balsa in test mode for the remaining 80 queries to gather the predicted cost and latency. 
% We then randomly split these data into calibration and test sets (JOB part). To evaluate its performance on a more complex distribution, we test Balsa on CEB~\cite{Negi2021FlowLoss}.

\noindent \textbf{Balsa.} \noindent\underline{\textit{Distribution Shift Quantification/Estimation.}} For validation of our adaptive {\cp} method, we first quantify the total variation distance of calibration distribution JOB ($\mathcal{D}_0$) and test distribution CEB ($\mathcal{D}$): $TV(\mathcal{D}, \mathcal{D}_0)$. Following the computation in Section~\ref{sec:cp_for_lqo_dynamic}, we randomly select \(500\) plans from JOB and CEB to empirically compute $tv:=TV(\mathcal{D},\mathcal{D}_0)= 0.0736$. We then set the allowed distribution shift $\epsilon = 0.08$ to ensure that $\epsilon$ exceeds the estimated distribution shift $tv$. 

% \begin{figure}[h]
%     \centering
%     \includegraphics[width=0.75\linewidth]{figures//evaluations/Balsa-TVDistance.png}
%     \caption{Balsa TV Distance}
%     \label{fig:balsa-tvdistance}
% \end{figure}
\noindent\underline{\textit{Validation Adaptive {\cp}}}. To maintain the original \((1-\delta)\) confidence level for the latency bounds, the adaptive {\cp} requires to obtain an adjusted upper bound \(\tilde{C}\) for the new distribution by computing an adjusted uncertainty probability $\tilde{\delta}$ with Equation~\ref{equ:tilde_delta_complex}. We set the uncertainty probability $\delta := 0.2$. We sample $K := 300$ calibration plans from JOB (\(\mathcal{D}_0 \)). We then compute the \(\text{Prob}(R^{(0)} \leq C(R^{(1)}, \dots, R^{(K)}))\) and \(\text{Prob}(R^{(0)} \leq \tilde{C}(R^{(1)}, \dots, R^{(K)}))\) for the non-adaptive and adaptive methods. Figure~\ref{fig:balsa-distribution-shift} shows the related results. In Figure~\ref{fig:adaptive-cp-balsa-c} (without performing adaptive {\cp}), the convergence is around 0.62, which is less than the expected \(1-\delta=0.8\). This shows the previously computed upper bound \(C\) was not suitable for the new distribution. However, in Figure~\ref{fig:adaptive-cp-balsa-adjusted-c} (with adaptive {\cp}), the coverage concentration is around 0.8, which shows our adjusted $\tilde{C}$ performs well with the new distribution.

% \begin{figure}[h]
%     \centering
%     \includegraphics[width=0.75\linewidth]{figures//evaluations/distribution_shift_c_and_adjusted_c.png}
%     \caption{Compare $C$ and $\tilde{C}$ within the calibration set (Balsa).}
%     \label{fig:distribution_shift_c_and_adjusted_c}
% \end{figure}

\begin{figure}[h]
    \centering\vspace{-10pt}
    \begin{subfigure}{0.48\linewidth}
        \centering
        \includegraphics[width=\linewidth]{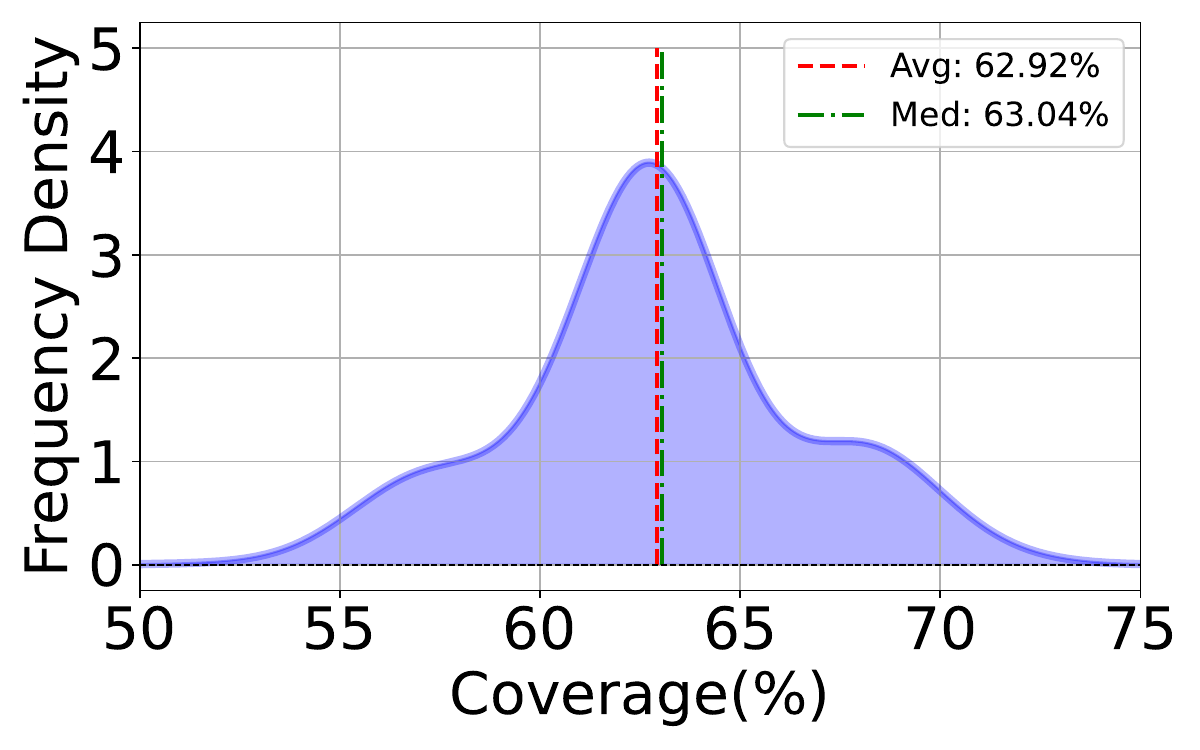}
        \vspace{-15pt}\caption{\(R^{(0)} \leq C(R^{(1)}, \dots, R^{(K)}))\)}
        \label{fig:adaptive-cp-balsa-c}
    \end{subfigure}
    \hfill
    \begin{subfigure}{0.48\linewidth}
        \centering
        \includegraphics[width=\linewidth]{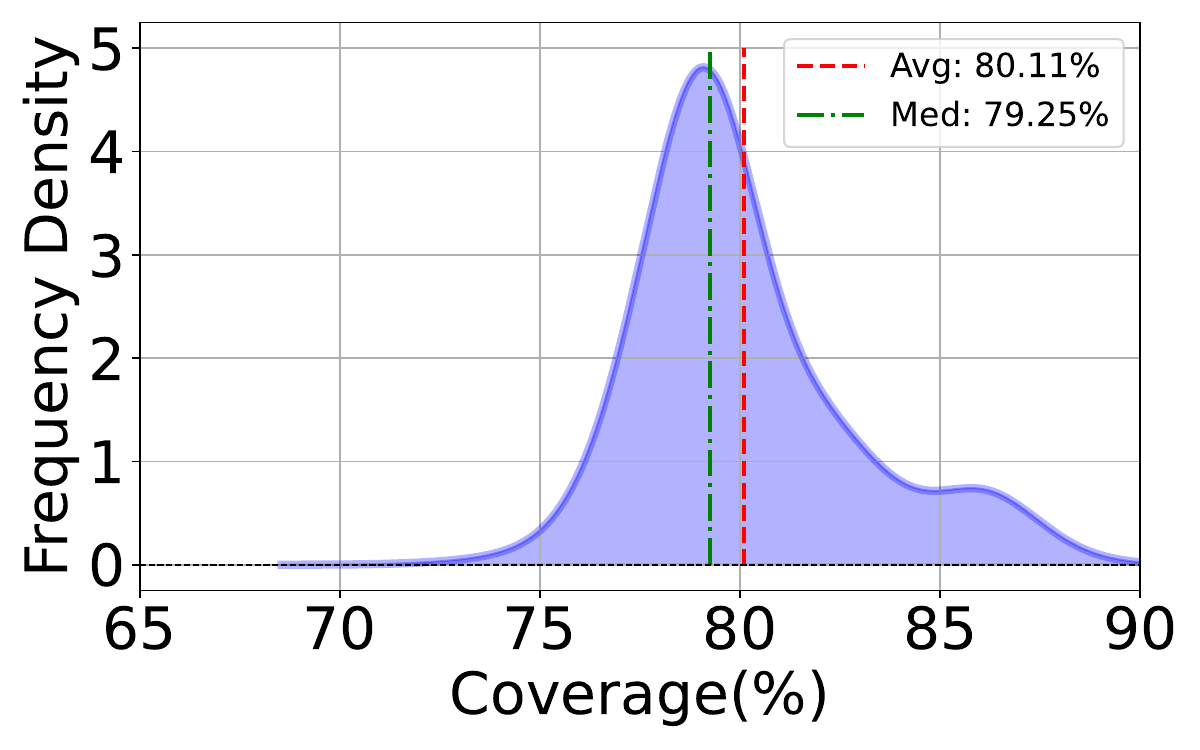}
        \vspace{-15pt}\caption{\(R^{(0)} \leq \tilde{C}(R^{(1)}, \dots, R^{(K)})\)}
        \label{fig:adaptive-cp-balsa-adjusted-c}
    \end{subfigure}
    \vspace{-8pt}\caption{JOB distribution shift with Adaptive {\cp} ($\delta=0.2$).}
    \label{fig:balsa-distribution-shift}
\end{figure}

% \begin{figure}
%     \centering
%     \includegraphics[width=0.75\linewidth]{experiments_results/balsa_0.2_distribution_shift.png}
%     \caption{JOB CEB Distribution Shift ($\delta=0.2$)}
%     \label{fig:enter-label}
% \end{figure}

% \subsubsection{RTOS}
\noindent \textbf{RTOS.}
We train RTOS on JOB ($\mathcal{D}_0$) and sample $K = 300$ plans to construct the calibration set. Then, we introduce a new distribution, JOB-light ($\mathcal{D}$), for testing. The TV distance between these two distributions is $tv = 0.24916$, then we set $\epsilon = 0.25$. Figure~\ref{fig:rtos-distribution-shift} shows the result for uncertainty levels $\delta=0.45$. The convergence of \(\text{Prob}(R^{(0)} \leq C(R^{(1)}, \dots, R^{(K)}))\) is around 0.2 and \(\text{Prob}(R^{(0)} \leq \tilde{C}(R^{(1)}, \dots, R^{(K)}))\) is 0.55 which is exactly \((1-\delta)\). This demonstrates that our adaptive {\cp} methods work well with different prototypes and different uncertainty conditions. 

% These results show that our adaptive {\cp} framework can handle different user uncertainty

% The final results for uncertainty levels $\delta = 0.4$ and $\delta = 0.5$ are presented in Figure~\ref{fig:rtos-distribution-shift}.

% \begin{figure}[h]
%     \centering
%     \begin{subfigure}{0.48\linewidth}
%         \centering
%         \includegraphics[width=\linewidth]{experiments_results/rtos_0.4_distribution_shift.png}
%         \caption{($\delta=0.4$}
%         \label{fig:rtos-0.4-shift}
%     \end{subfigure}
%     \hfill
%     \begin{subfigure}{0.48\linewidth}
%         \centering
%         \includegraphics[width=\linewidth]{experiments_results/RTOS_0.5_distribution_shift.png}
%         \caption{$\delta=0.5$}
%         \label{fig:rtos-0.5-shift}
%     \end{subfigure}
%     \caption{RTOS distribution shift with different uncertainty levels. (I can make a fine-grained graph to make it look nicer)}
%     \label{fig:rtos-distribution-shift}
% \end{figure}

\begin{figure}[h]
    \centering\vspace{-10pt}
    \begin{subfigure}{0.48\linewidth}
        \centering
        \includegraphics[width=\linewidth]{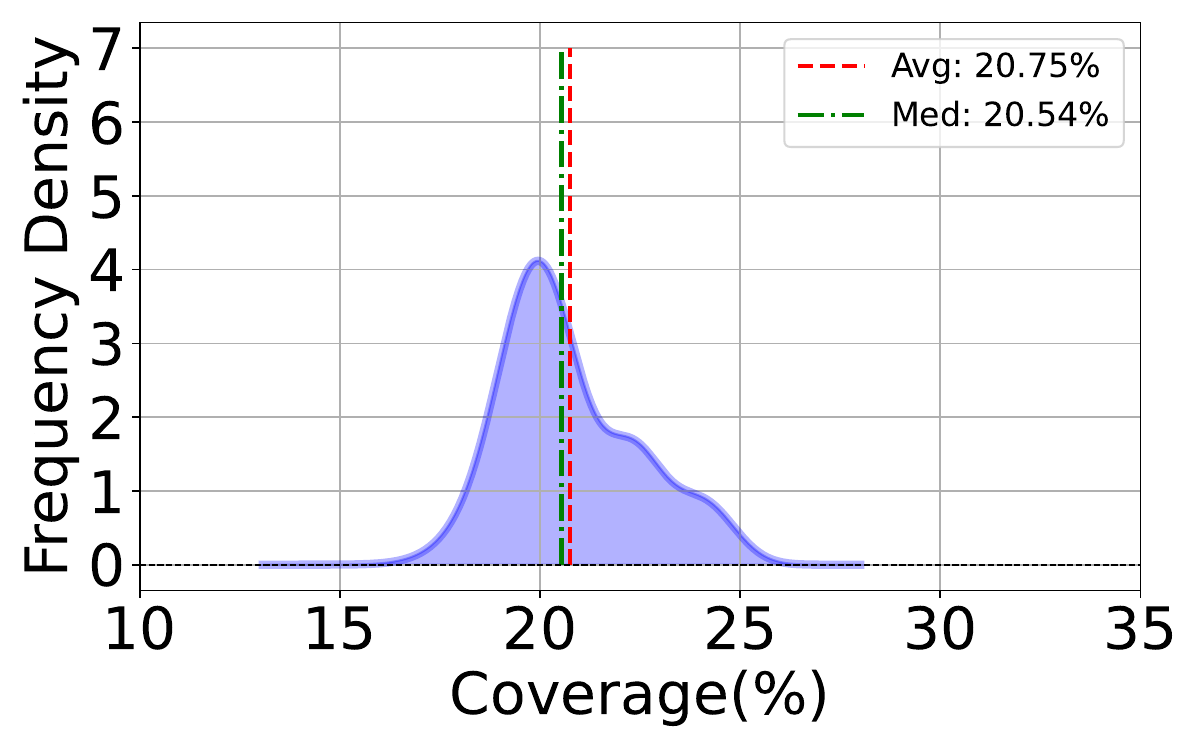}
        \vspace{-15pt}\caption{\(R^{(0)} \leq C(R^{(1)}, \dots, R^{(K)}))\)}
        \label{fig:adaptive-cp-rtos-c}
    \end{subfigure}
    \hfill
    \begin{subfigure}{0.48\linewidth}
        \centering
        \includegraphics[width=\linewidth]{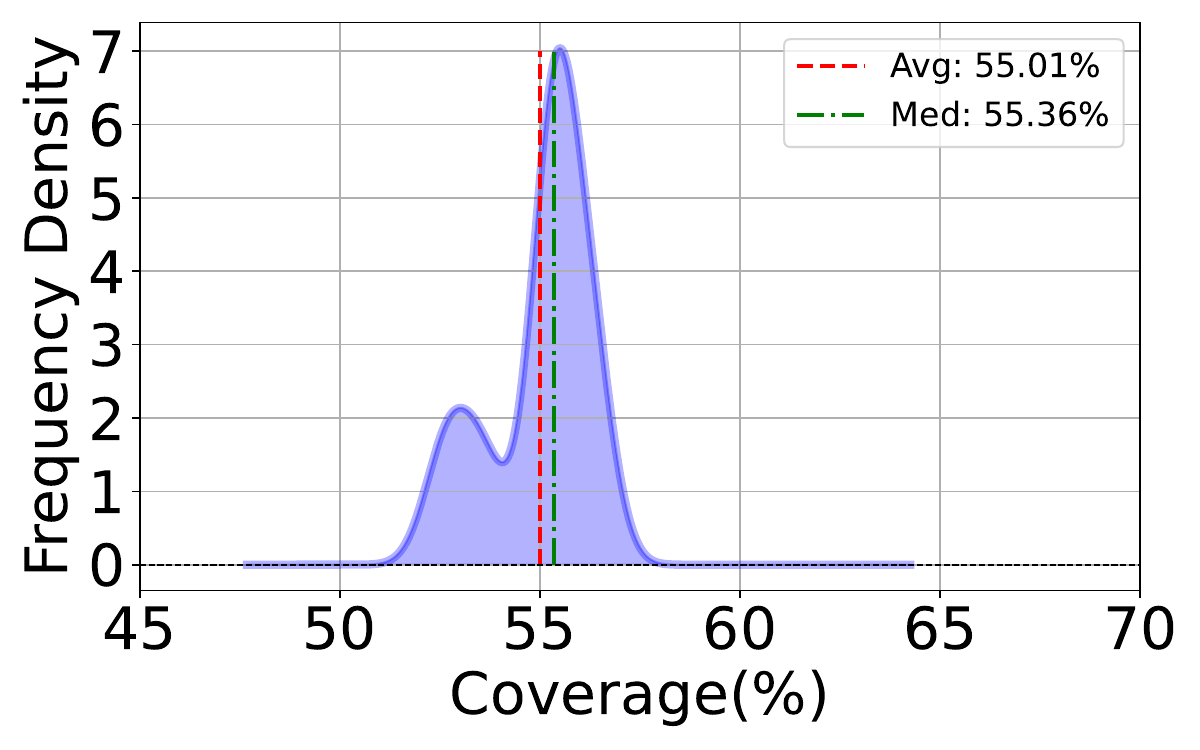}
        \vspace{-15pt}\caption{\(R^{(0)} \leq \tilde{C}(R^{(1)}, \dots, R^{(K)})\)}
        \label{fig:adaptive-cp-rtos-adjusted-c}
    \end{subfigure}
    \vspace{-10pt}\caption{RTOS distribution shift with Adaptive {\cp} ($\delta=0.45$).}
    \label{fig:rtos-distribution-shift}
\vspace{-10pt}\end{figure}
% We can observe the following:  
% 1) For $\delta = 0.4$, the new coverage [[update]] on the new distribution is above $1 - \delta = 0.6$.  
% 2) For $\delta = 0.5$, the median coverage across 10,000 iterations is 0.55[[double check]], which also demonstrates that our adjusted $\tilde{C}$ performs well with the new distribution under different uncertainty setups. These results confirm the robustness of our theorem.
}
\vspace{-6pt}\subsection{Runtime Verification}\label{sec:eval-runtime-verification}
We perform our runtime verification evaluation with {\balsa} as a white-box {\lqo} \revised{using the JOB workload}. In this experiment, we aim to validate Lemma~2. Specifically, we aim to demonstrate the following statement:
\[
P(X \models \phi) \geq 1 - \delta, \quad \text{if} \quad \rho^\phi(\hat{x}) > C
\]
We define our performance constraint with the following STL specification:
\[
\phi := G(X < threshold)
\]
where \(G\) is the \textit{always} operator defined in Section~\ref{sec:background_cp}. We use this specification to bound the actual latency \(X\) when running an {\lqo}'s plan, whether partial or complete: the value of \(X\) is expected to always be less than the threshold. We set the threshold as 1000 and 2000, which implies that the cumulative latency of operations should not exceed $1000ms$ and $2000ms$ in the database context. We use this STL to detect violations and avoid unexpected long latency in execution. Based on $\phi$, we define the robust semantics as follows:
\[
\rho^{\phi} (x) = threshold - x
\]

\begin{figure}[h]
    \centering\vspace{-10pt}
    \begin{subfigure}{0.48\linewidth}
        \centering
        \includegraphics[width=\linewidth]{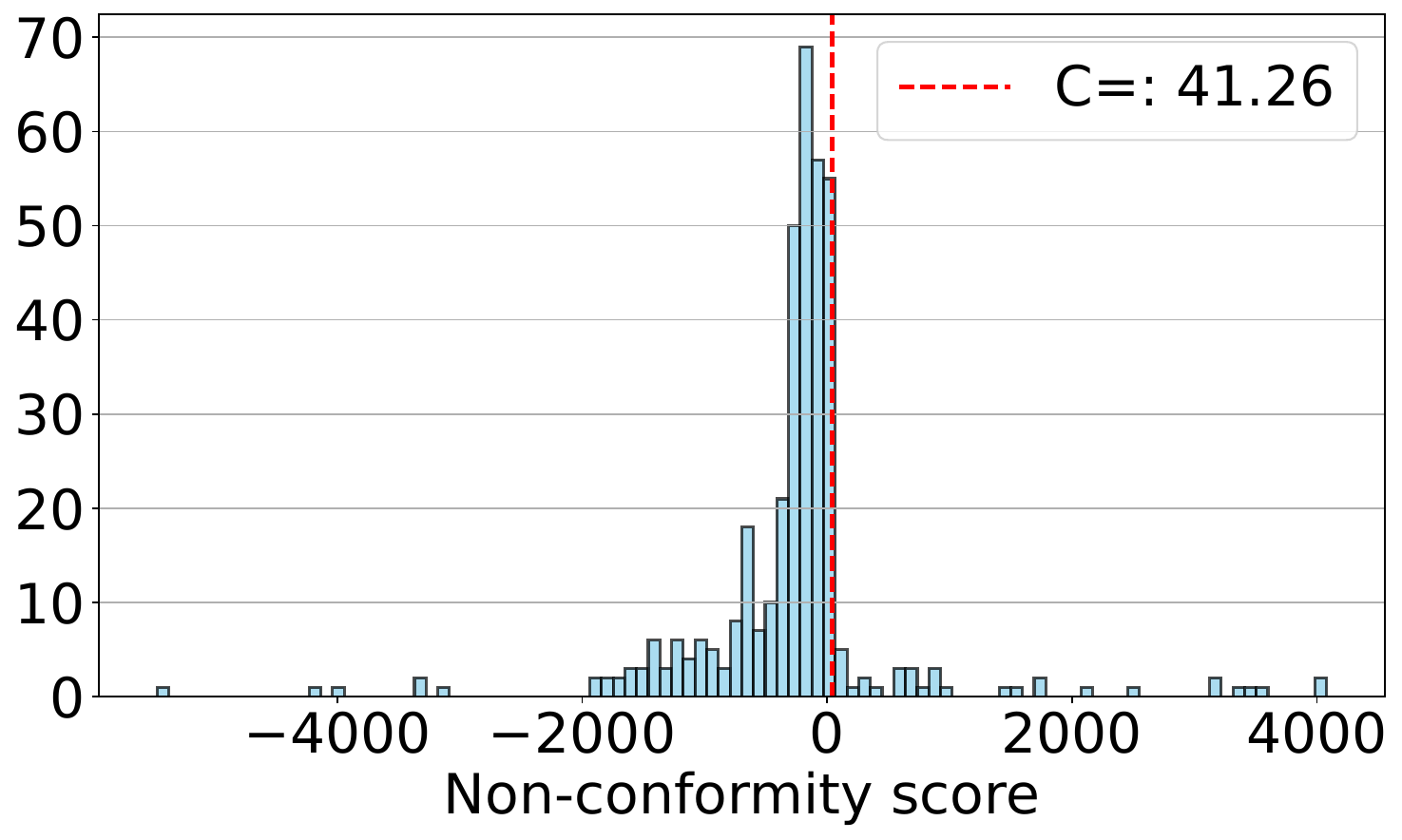}
        \vspace{-15pt}\caption{Threshold = 1000}
        \label{fig:RuntimeVerification1000}
    \end{subfigure}
    \hfill
    \begin{subfigure}{0.48\linewidth}
        \centering
        \includegraphics[width=\linewidth]{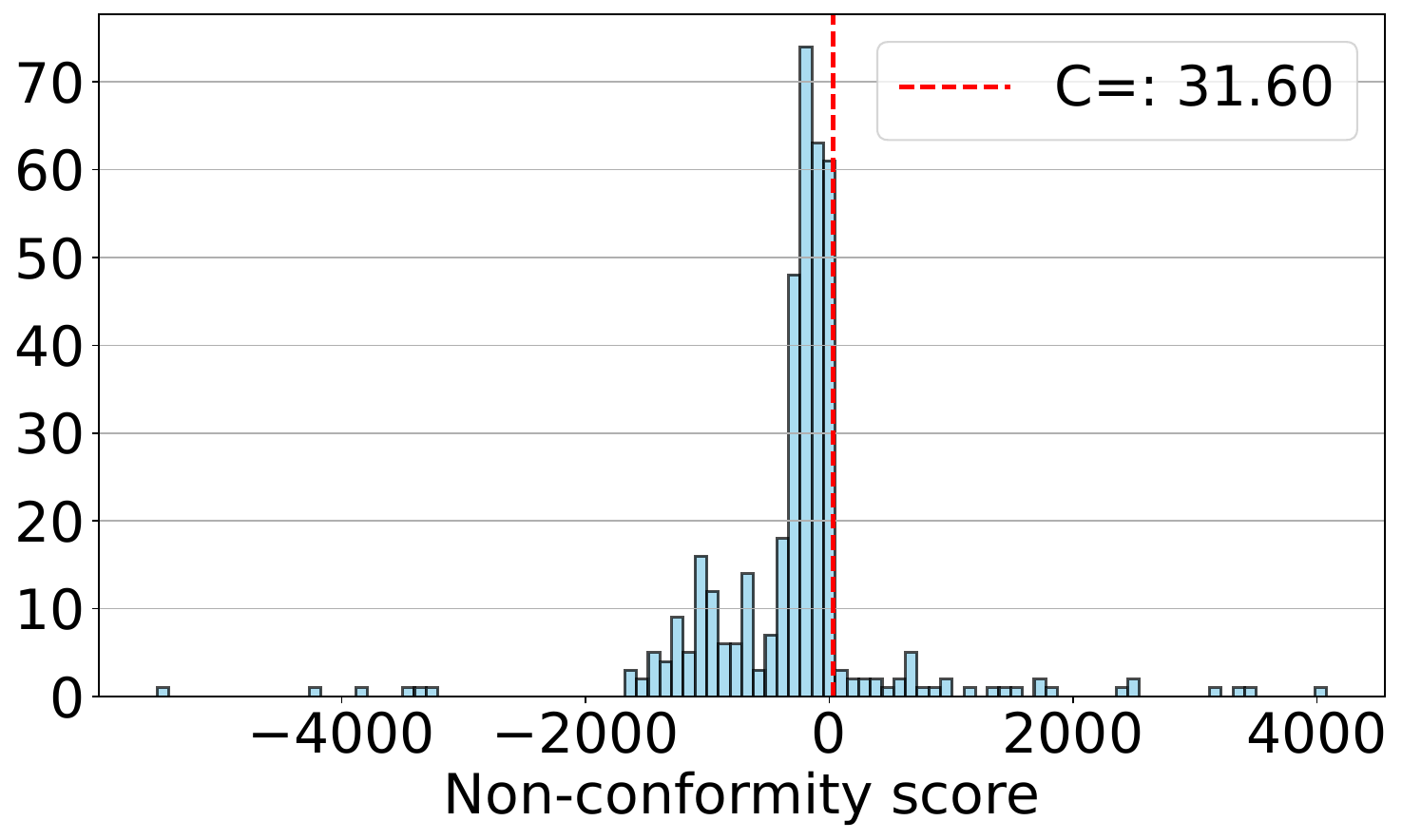}
        \vspace{-15pt}\caption{Threshold = 2000}
        \label{fig:RuntimeVerification2000}
    \end{subfigure}
    \vspace{-10pt}\caption{Non-conformity Scores in Runtime Verification.}
    \label{fig:eval-nonconformity-score}
\vspace{-10pt}\end{figure}

From the calibration queries, we construct the value of \(C\) and use this value to verify whether \(\rho^\phi(\hat{x}) > C\). If this holds true, the actual latency adheres to the STL specification. Figure~\ref{fig:eval-nonconformity-score} shows the non-conformity score distribution with different thresholds. Our unified-based upper bound $C$ covers $1 - \delta = 90\%$ of the non-conformity scores (left side of the red dashed line).

\noindent{\boldmath{$\phi := G(X < 1000)$}:} We found that for 27 of the 30 queries (\( \lvert Q^{\text{test}} \rvert = 30 \)), it holds that \(\rho^\phi(\hat{x}) > C\) implies \(X \models \phi\), confirming the correctness of runtime verification (Lemma 2). We also validated Equation~\ref{3.2-2} and found that 28 of the 30 test queries satisfy \(\rho^\phi(\hat{x}) - \rho^\phi(X) \leq C\), which is greater than $(1 - \delta) = 0.9$, further confirming the correctness of {\cp}.

\noindent{\boldmath{$\phi := G(X < 2000)$}:} This is a looser threshold. We found that for 29 of the 30 queries (\( \lvert Q^{\text{test}} \rvert = 30 \)), it holds that \(\rho^\phi(\hat{x}) > C\) implies \(X \models \phi\). A larger threshold demonstrates better coverage. We also validated Equation~\ref{3.2-2} and found that 29 of the 30 test queries satisfy \(\rho^\phi(\hat{x}) - \rho^\phi(X) \leq C\), further confirming our method.

% OCT-29 
% \begin{figure}
%     \centering
%     \includegraphics[width=0.9\linewidth]{figures//1000iter//BalsaXImdb/runtimeVerificationR.png}
%     \caption{Runtime verification with threshold 2000}
%     \label{fig:enter-label}
% \end{figure}
% \begin{figure}
%     \centering
%     \includegraphics[width=0.9\linewidth]{figures//1000iter//BalsaXImdb/RuntimeVerificationR1000.png}
%     \caption{Runtime verification with threshold 1000}
%     \label{fig:enter-label}
% \end{figure}

\vspace{-6pt}\subsection{Violation Detection and Handling}\label{sec:eval-violation-detection-and-handling}
% \insights{We notice that for queries that are flagged, postgres generally performs better. There are some cases where postgres plans are worse than balsa's plans but over all the queries the latency savings for these flagged queries are considerable.}
\revised{We perform violation detection using the JOB workload as discussed in Lemma 2 over the constraint ($\phi := G(X < 2000)$).} If violations are detected, we introduce PostgreSQL to assist in generating a new query plan for execution. We compare two scenarios: \textit{with {\cp}} and \textit{without {\cp}}, representing {\cp}-based violation detection and normal {\lqo} planning, respectively. In this section, we focus on comparing the plan quality between these two methods.

\begin{figure}[htbp]
\centering \vspace{-10pt}
\includegraphics[width=0.9\linewidth]{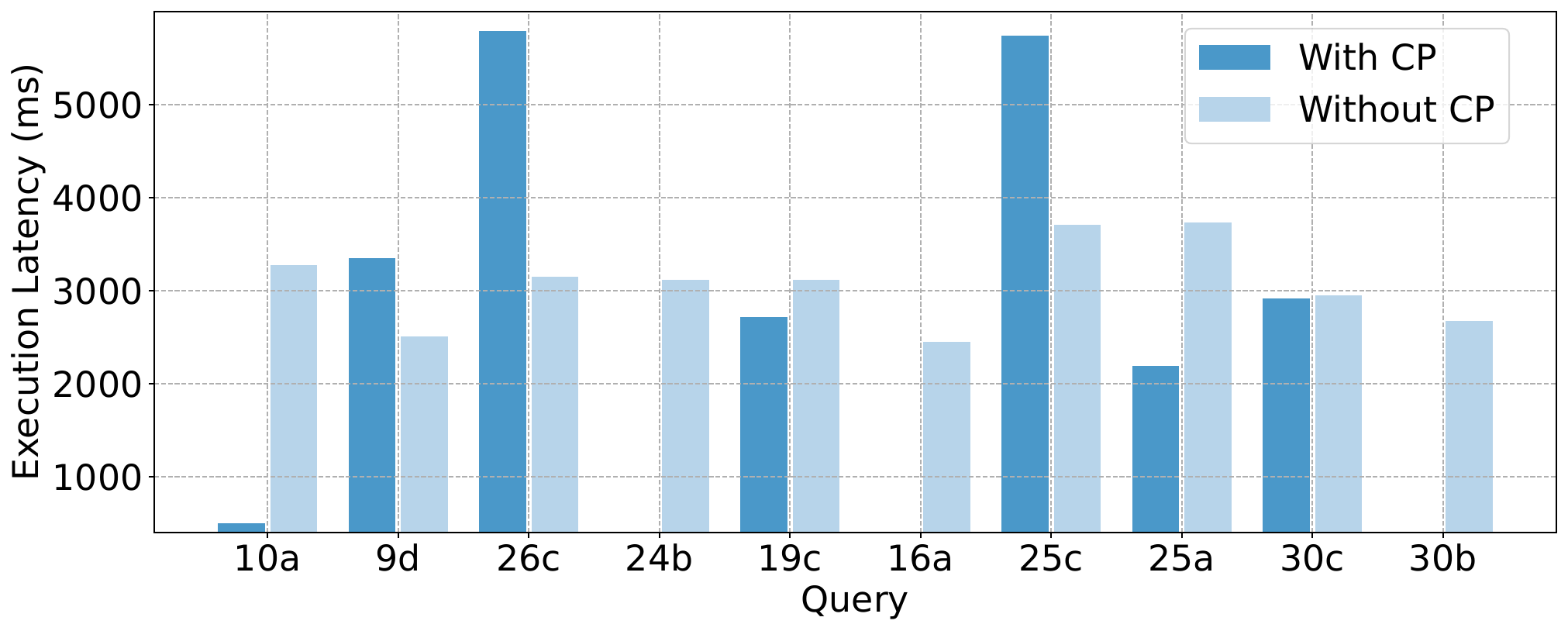}
\vspace{-8pt}\caption{Violation Detection \revised{($\phi := G(X < 2000)$)}: Latency Comparison With and Without {\cp} (Balsa).} 
\label{fig:ViolationDetectionBalsa}
\vspace{-10pt}\end{figure}

\noindent{\textbf{Balsa with Violation Detection.}}
Figure~\ref{fig:ViolationDetectionBalsa} presents the comparison results for Balsa. In total, 10 queries were flagged as potential violations. We trigger PostgreSQL to re-generate the query plans. Notably, for 7 out of these 10 queries, the query plans generated by PostgreSQL outperformed the Balsa-generated plans. The overall latency savings for these 7 queries amounted to 22.12\%. \revised{For the remaining 3 queries, we observed that although {\balsa} results in better plans for them, these plans still violate the user constraint. That is why these queries are still detected by our verification framework.}

% The overall latency savings for these 7 queries amounted to \textit{12,030.1 ms}.

% \tohanwen{check queries => describe}

\vspace{-6pt}\subsection{{\cp}-Guided Actual Latency Upper Bound Query Optimizer}\label{sec:eval-cp-guided}

In this section, we still use Balsa as a representative white-box {\lqo} to conduct {\cp}-guided plan search experiments. 
% \sout{We integrated our proposed Beam Search variation in Section~\ref{sec:cp_guided} into {\balsa}.} 
\revised{Considering Balsa uses beam search~\cite{lowerre1976harpy} internally, our discussion revolves around {\cp}-guided beam search.} We evaluated Balsa at different training epochs: 50, 100, and 150, corresponding to moderately trained, well-trained, and highly trained Balsa, respectively. To evaluate our method, we use 33 queries from template \textit{b} as the test set and the other 47 queries as the calibration set. 
% set of 47 queries, which includes 31 "*c.sql" files, 12 "*d.sql" files, 2 "*e.sql" files, and 2 "*f.sql" files. 
The comparison experiments are conducted five times, and the average is reported to reduce the impact of system fluctuations on the planning and execution time.

\subsubsection{Plan Improvement}

\begin{figure}[h]
    \centering\vspace{-10pt}
    \includegraphics[width=0.9\linewidth]{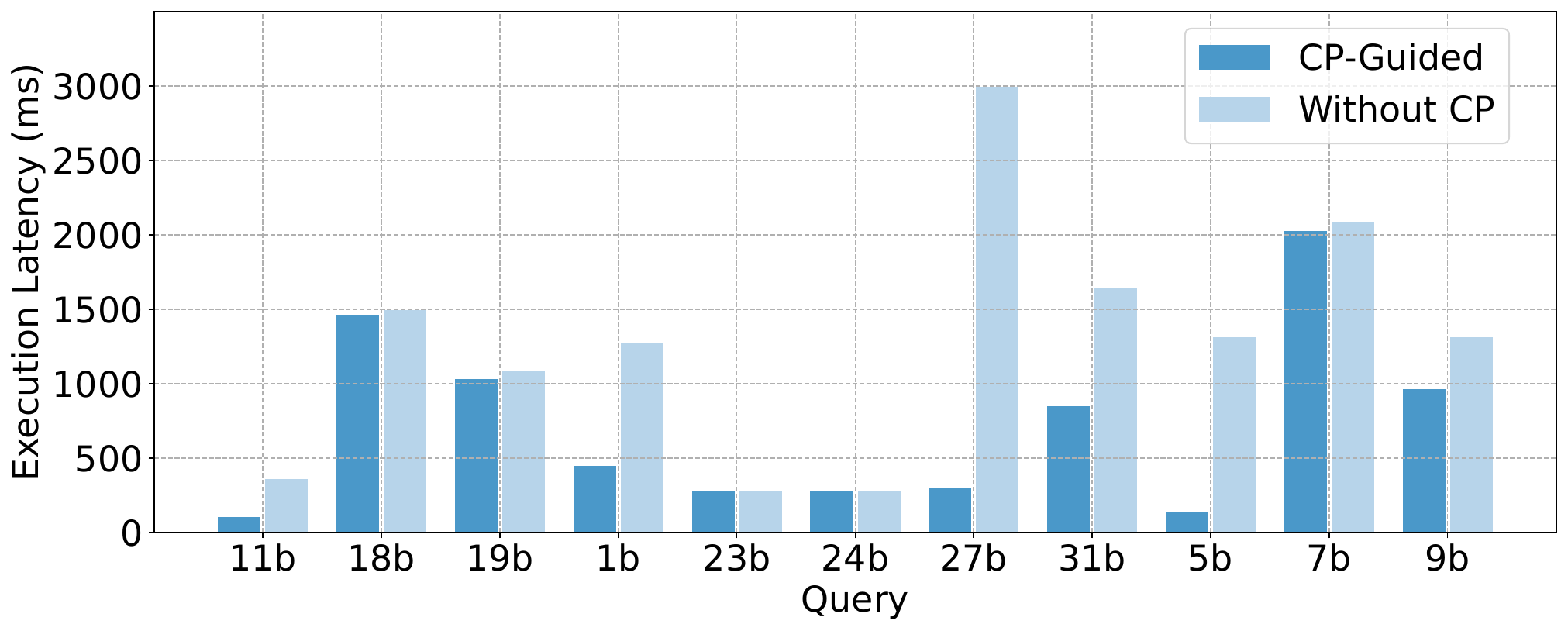}
    \vspace{-14pt}\caption{Plan Quality Comparison: CP Guided Algorithm vs. Balsa (50 iterations)}
    \label{fig:cp-guided-50-iterations}
\vspace{-10pt}\end{figure}

Using the {\cp}-guided plan search, we employ the {\cp}-guaranteed latency upper bound as a heuristic to guide the beam search in constructing complete query plans. We evaluate whether this approach yields better results compared to the vanilla Balsa. Figure~\ref{fig:cp-guided-50-iterations} shows the queries where we achieve improvements, with plan enhancements observed in 11 out of 33  test queries while the rest maintained the same plan quality. This demonstrates that our algorithm can effectively improve plan quality for a {\lqo}. 

\begin{figure}[h]
    \centering\vspace{-10pt}
    \begin{subfigure}[b]{0.45\linewidth}
        \centering
        \includegraphics[width=\linewidth]{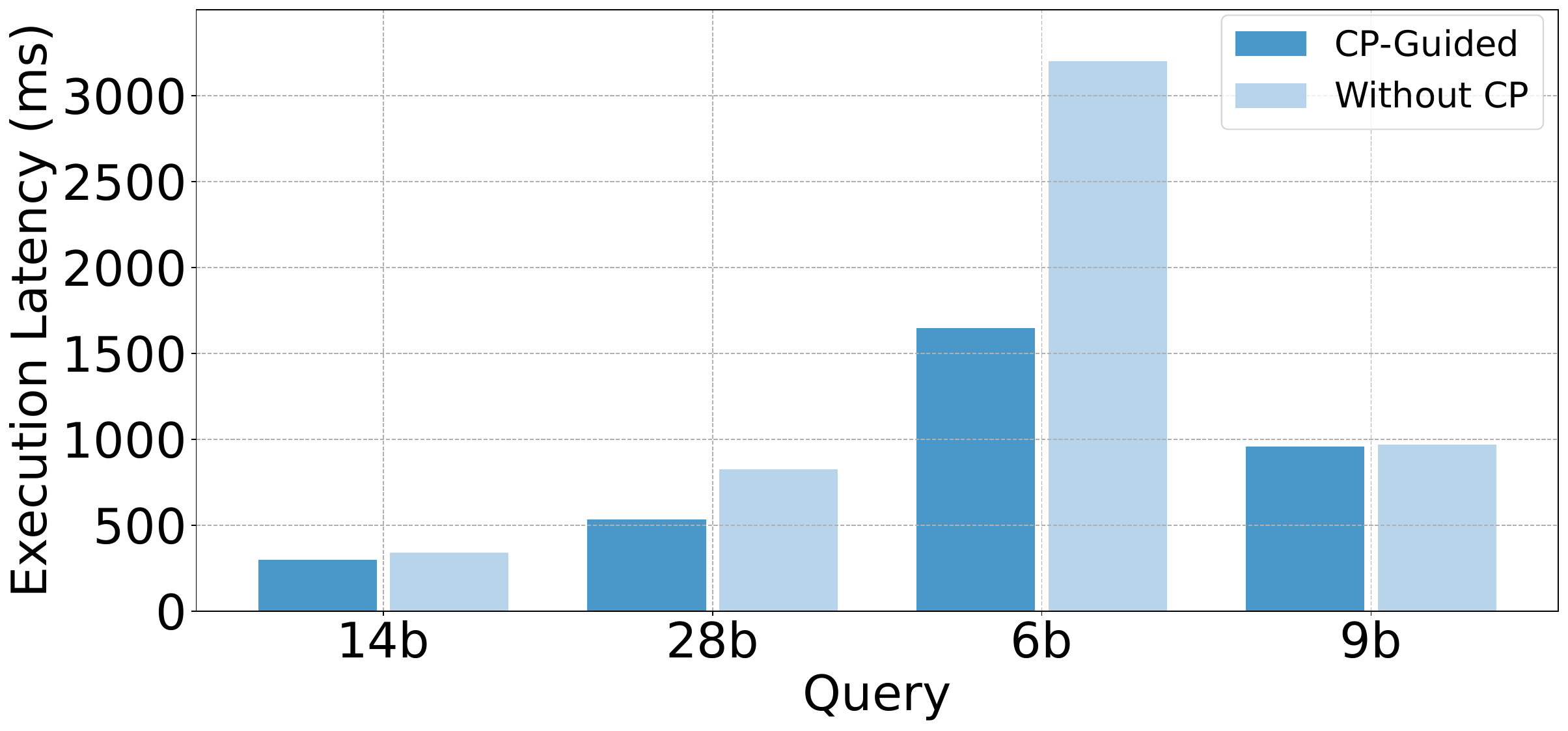}
        \vspace{-14pt}\caption{Balsa, 100 iterations}
        \label{fig:cp-guided-100-iterations}
    \end{subfigure}
    \hfill
    \begin{subfigure}[b]{0.45\linewidth}
        \centering
        \includegraphics[width=\linewidth]{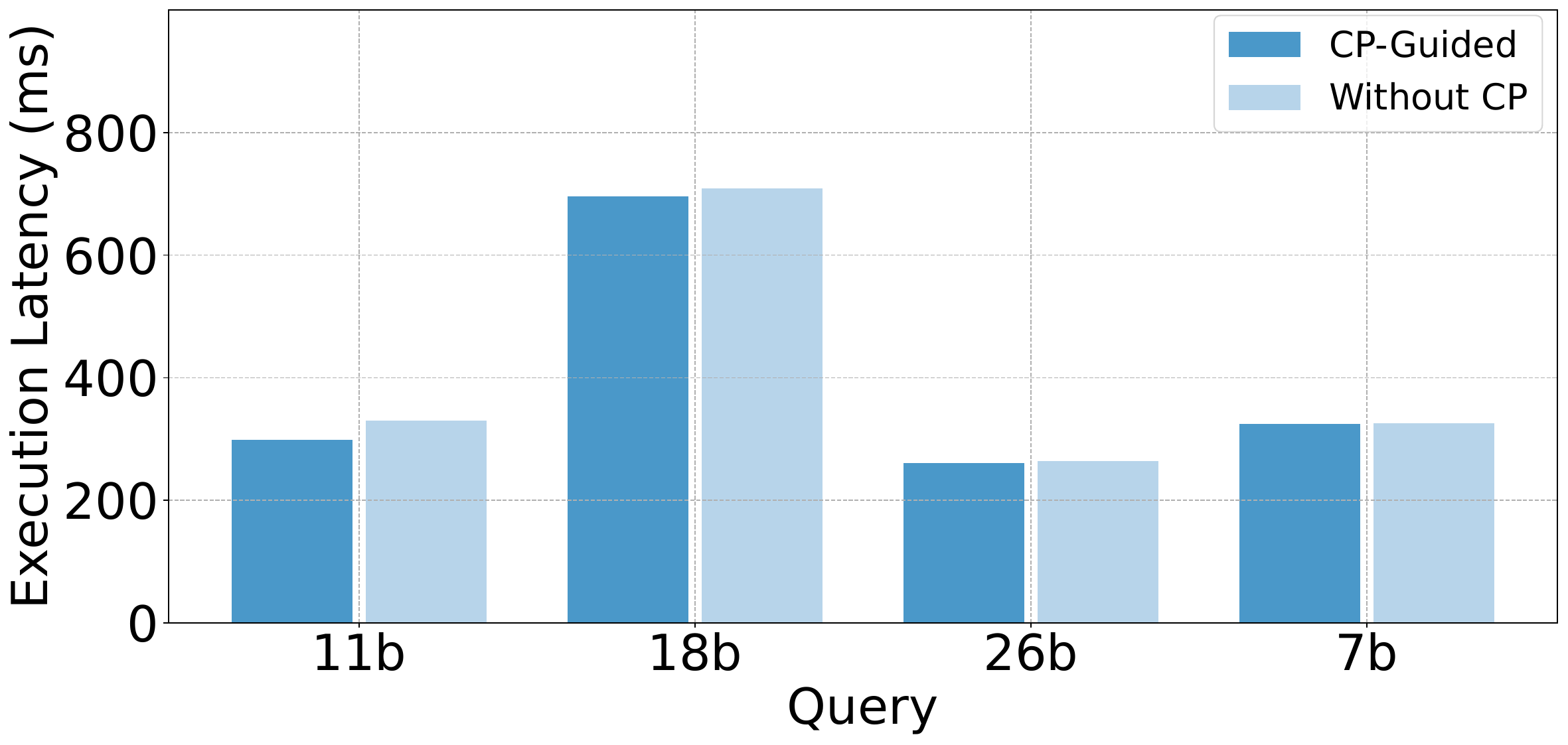}
        \vspace{-14pt}\caption{Balsa, 150 iterations}
        \label{fig:cp-guided-150-iterations}
    \end{subfigure}
    \vspace{-10pt}\caption{Plan Quality Comparison.}
    \label{fig:cp-guided-comparison}
\vspace{-10pt}\end{figure}

For a well-trained Balsa (100 iterations), our algorithm improves the plan quality for queries 14b, 28b, 6b and 9b as seen in Figure \ref{fig:cp-guided-100-iterations}, demonstrating consistent plan improvement. Even for the highly trained Balsa (150 iterations), we also observe several improved queries as seen in Figure \ref{fig:cp-guided-150-iterations}. Although Balsa can reliably and efficiently generate high-quality query plans at this stage, the {\cp}-guided algorithm can still achieve better plans, even within this highly constrained search space. This further proves the effectiveness of our algorithm. 

We also observe that our algorithm achieves greater improvements in plan quality during the early training stages of Balsa. This aligns with the intuition that it is easier to make improvements within a larger discovery space. As the number of training iterations increases, Balsa becomes progressively more refined, which naturally narrows the scope for further improvement.

We perform a deep-dive analysis of the queries where we achieve significant improvements: Query 6b in Figure~\ref{fig:cp-guided-comparison}~(a) and Query 27b in Figure~\ref{fig:cp-guided-50-iterations}. When we closely compare the query plans generated by {\cp}-Guided and those without {\cp} guidance, we observe that in Query 6b, Balsa originally selects a pattern of (NL, NL, IS). However, in the {\cp}-guided plan search algorithm, we instead select a pattern of (HJ, NL, SS). The (HJ, NL, SS) pattern aligns with the valid patterns established for our reliable {\cp} construction, whereas (NL, NL, IS) is not among them. By following our algorithm and being guided by {\cp}, Query 6b achieves 48.52\% latency reduction by replacing this pattern. For Query 27b, our {\cp}-guided approach has an even greater impact. Without {\cp} guidance, Balsa generates a left-deep tree; however, under {\cp} guidance, it produces a bushy tree, resulting in a 9.84x improvement in latency. Query-level analysis reveals that our algorithm not only favors reliable patterns to construct the entire query plan but can also systematically optimize the structure of query plan, significantly enhancing the overall query plan quality.

\subsubsection{Planning Time Comparison}
\begin{figure}[h]
    \centering\vspace{-10pt}
    \includegraphics[width=0.9\linewidth]{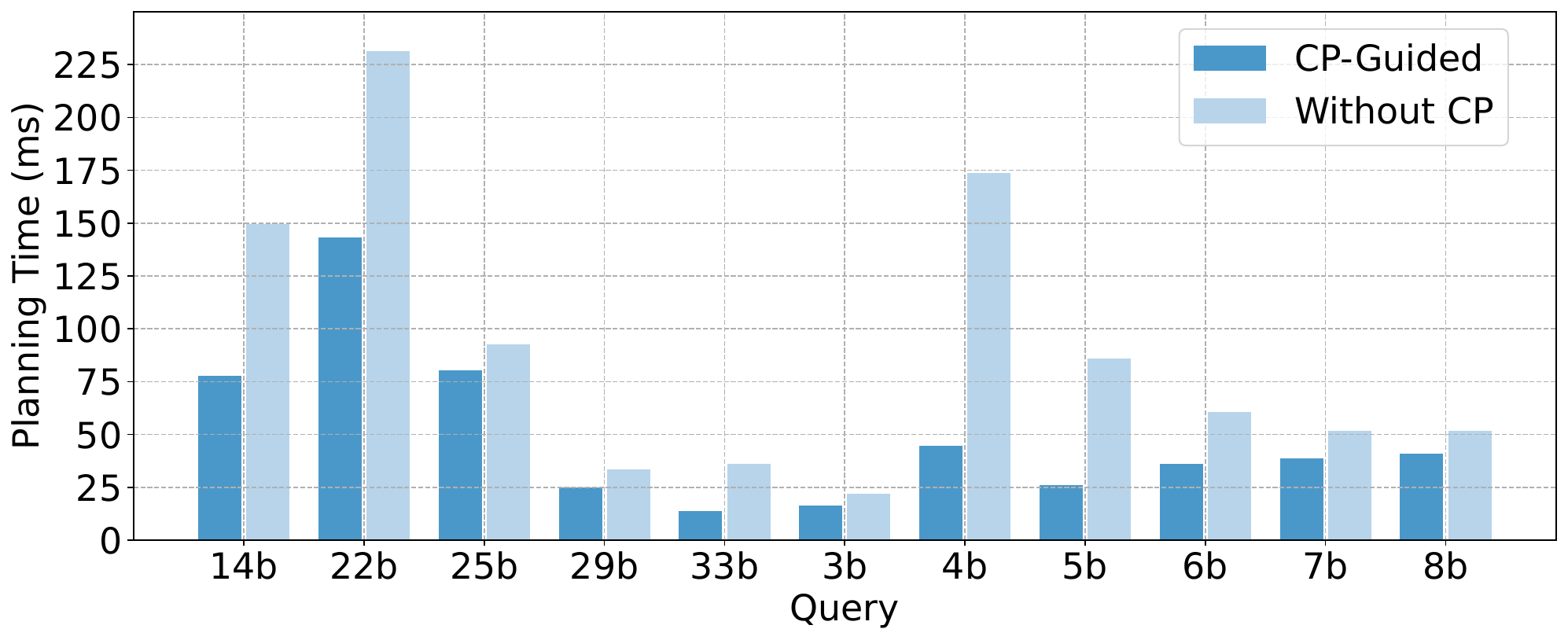}
    \vspace{-14pt}\caption{Planning Time Comparison: CP Guided Algorithm vs. Balsa (50 iterations)}
    \label{fig:planning time-50 iterations}\vspace{-10pt}
\end{figure}

For a moderately trained Balsa, we observe an improvement in planning time. Without {\cp} assistance, the total planning time for all test queries is \textit{6178.60 ms}; however, with our {\cp}-guided algorithm, it is reduced to \textit{5563.40 ms}, achieving an overall improvement of 9.96\%. This demonstrates that our {\cp}-guided approach can mitigate suboptimal {\lqo} behaviors and accelerate the plan search. For the single query level, we can also observe Query 4b in Figure \ref{fig:planning time-50 iterations} reduces 74.40\% planning time. This effect can be attributed to the optimization target of the {\cp}-guided algorithm—the actual latency upper bound—which acts as a stricter heuristic than previously cost itself. This leads to a more direct search path within the search space. Compared to a moderately trained Balsa, our algorithm constrains the search scope, thereby reducing planning time.

\begin{figure}[h]
    \centering
    \vspace{-10pt}
    \includegraphics[width=0.9\linewidth]{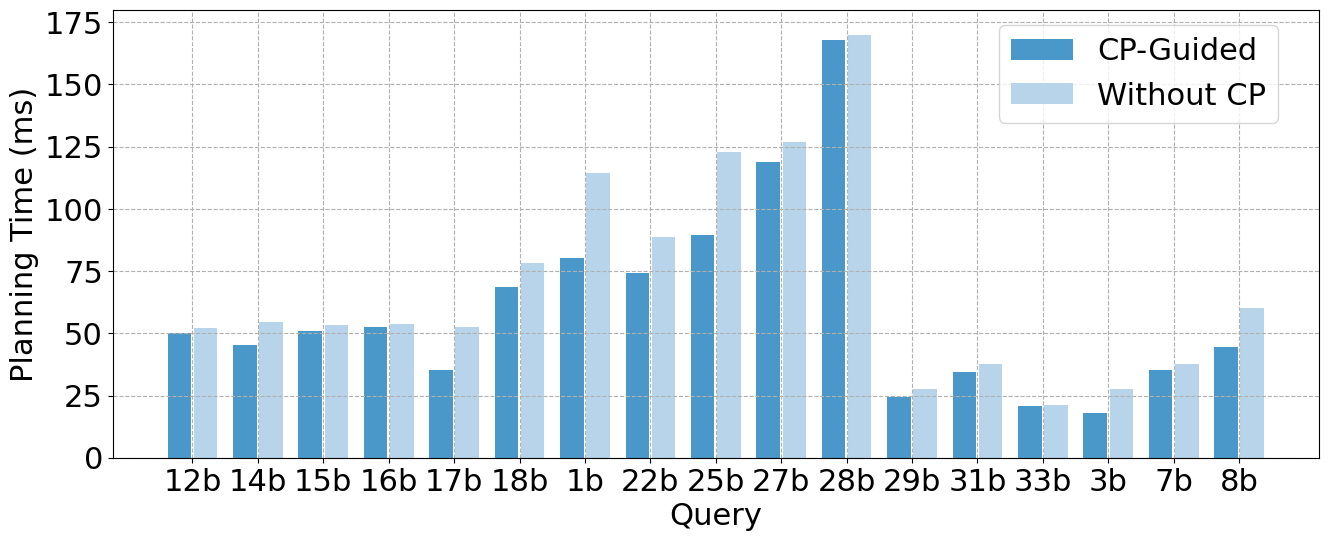}
    \vspace{-8pt}\caption{Planning Time Comparison: CP Guided Algorithm vs. Balsa (150 iterations)}
    \label{fig:planning time-150 iterations}\vspace{-10pt}
\end{figure}

Figure~\ref{fig:planning time-150 iterations} illustrates that even with a highly trained Balsa, our algorithm improves planning time for 17 out of 33 queries. We also observe that as the number of {\lqo} training iterations increases, the overall planning time for both {\cp}-guided and without {\cp} methods decreases. Comparing Figure~\ref{fig:planning time-50 iterations} and Figure~\ref{fig:planning time-150 iterations}, we can see that the impact of our {\cp}-guided algorithm on planning time is more pronounced at lower training iterations. This is because, with more extensive training, the {\lqo} has a more refined initial search direction, resulting in a relatively smaller search space for our algorithm. 
Notably, for Queries 7b and 18b, we achieve improvements in both plan quality and planning time. These observations further demonstrate the effectiveness of our {\cp}-guided algorithm.

\vspace{-10pt}\revised{\subsection{Hyper-Parameter Micro-benchmarking.}
\label{sec:eval-hyper-parameter}}

In this section, We discuss \revised{three} types of hyper-parameters and observe their impact on the coverage.

\begin{figure}[h]
    \centering
    \vspace{-10pt}
    \begin{subfigure}{0.48\linewidth}
        \centering
        \includegraphics[width=\linewidth]{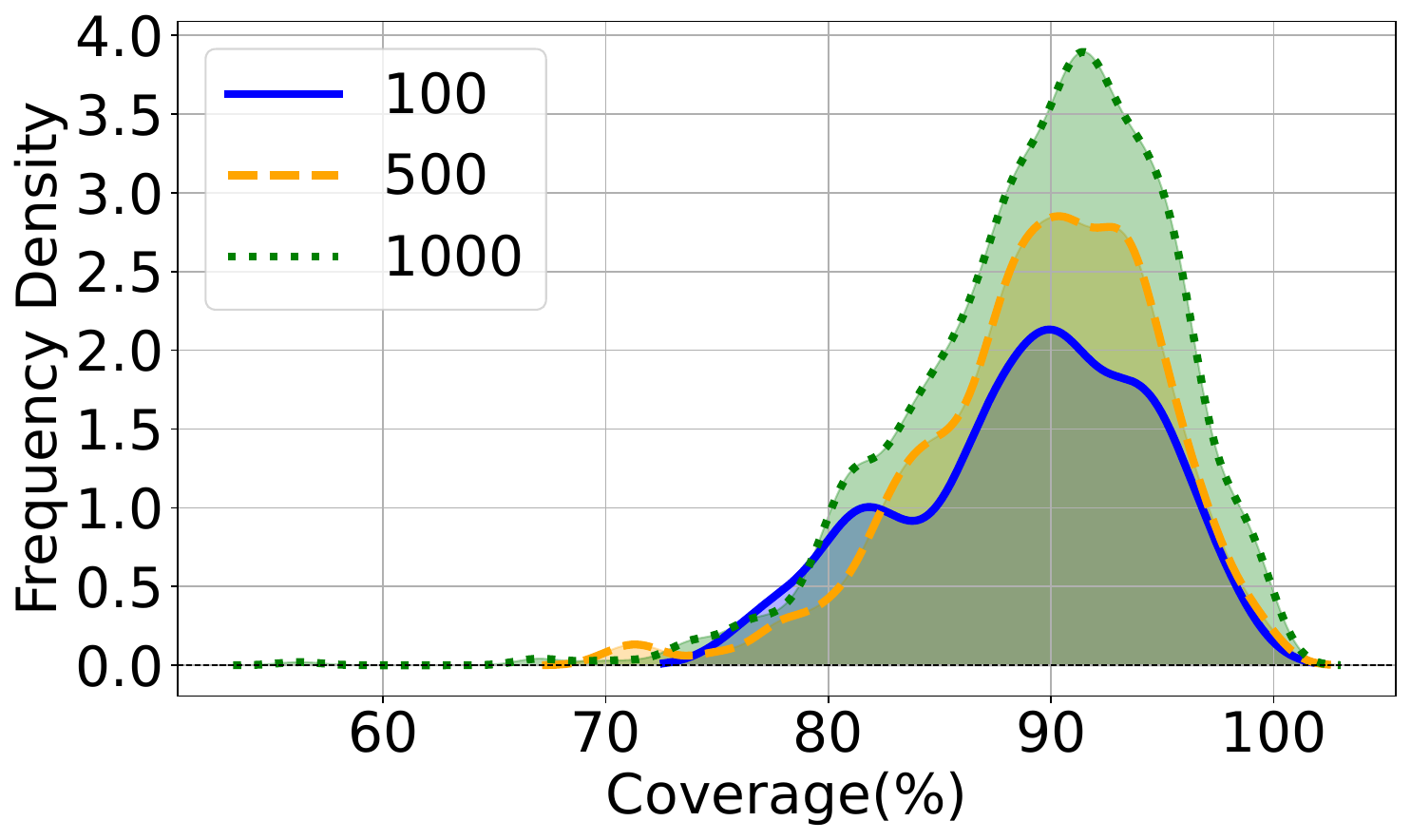}
        \vspace{-15pt}\caption{Balsa on JOB}
        \label{fig:impact-sampling-job}
    \end{subfigure}
    \hfill
    \begin{subfigure}{0.48\linewidth}
        \centering
        \includegraphics[width=\linewidth]{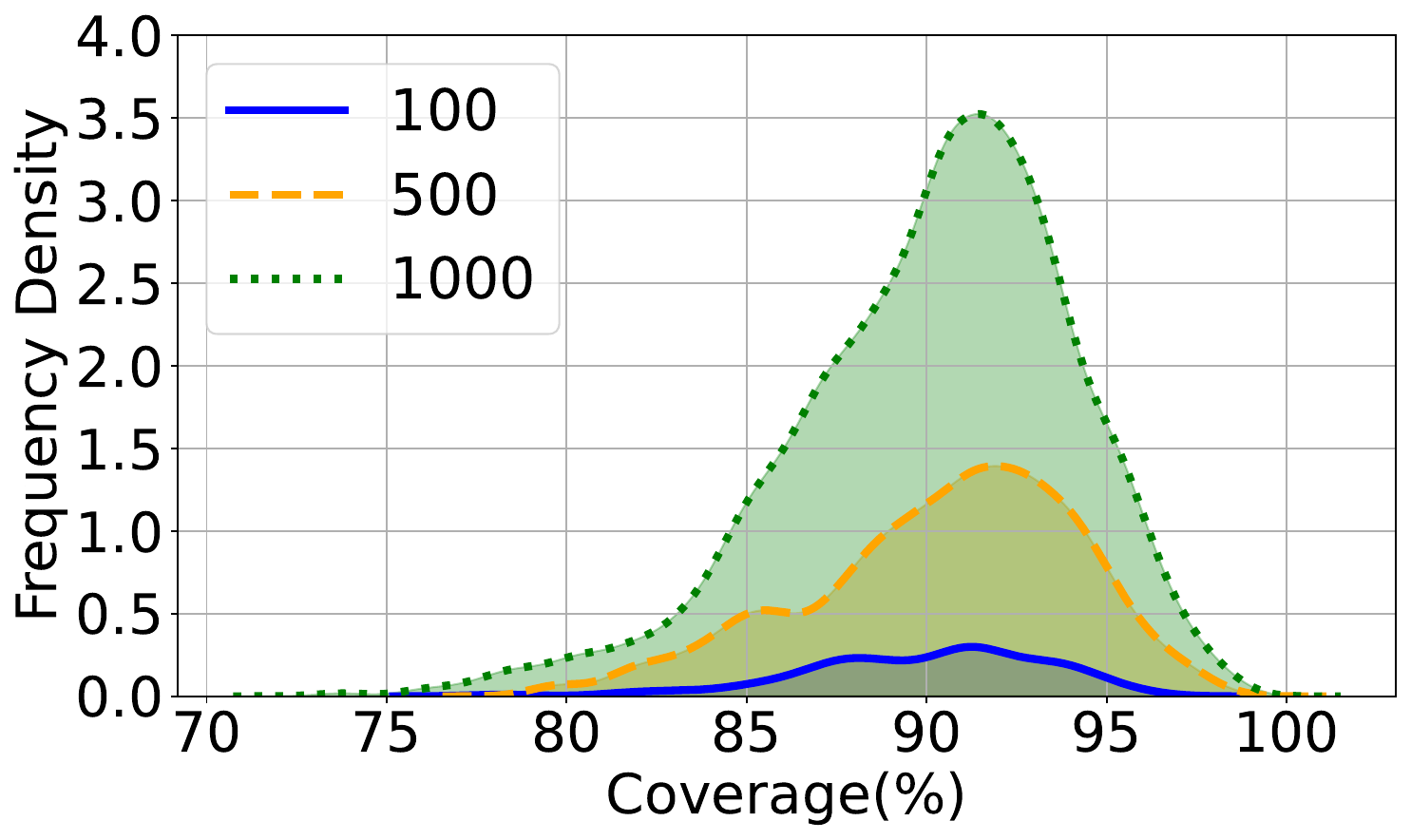}
        \vspace{-15pt}\caption{Balsa on TPC-H}
        \label{fig:impact-sampling-tpch}
    \end{subfigure}
    \vspace{-10pt}\caption{Impact of Changing the Sampling Iterations.}
    \label{fig:impact-sampling}
\vspace{-10pt}\end{figure}

\noindent{\textbf{Impact of Changing the Sampling Iterations.}}
% \subsubsection{Impact of Changing the Sampling Iterations}
We begin by examining how the first hyper-parameter—sampling iterations—affects empirical coverage. We test with 100, 500, and 1000 sampling iterations. For each sampling iteration setting, we plot the density of each coverage. Figure~\ref{fig:impact-sampling}~(a) and Figure~\ref{fig:impact-sampling}~(b) illustrate Balsa's performance on the {\job} and {\tpch} workloads, respectively. When the number of sampling iterations is low, the curve appears less smooth due to limited sampling. Since empirical coverage approximates the inherent coverage properties of {\cp}, insufficient sampling fails to capture the expected behavior according to {\cp} theory. With more iterations, the curve smooths, more accurately reflecting the intrinsic coverage properties of {\cp} theory. Additionally, the curve displays a sharper peak shape. %In all the following experiments, we always set sampling iterations as 1000.

% \sout{We set $\delta = 0.1$. According to Equation~\ref{4.3-2}, the {\cp} theory predicts that the most frequent coverage should be greater than $1 - \delta = 0.9$, as reflected by the peak of the curve in both graphs. For both workloads, the peak of the green dashed line demonstrates this trend, empirically validating the correctness of applying {\cp} with {\lqo}s.} 
We also observe that the JOB workload exhibits a higher frequency density than the TPC-H workload. This is because JOB contains more joins, leading to a greater number of validation data points, which increases the frequency density.

\begin{figure}[h]
    \centering
    \vspace{-10pt}
    \begin{subfigure}{0.48\linewidth}
        \centering
        \includegraphics[width=0.8\linewidth]{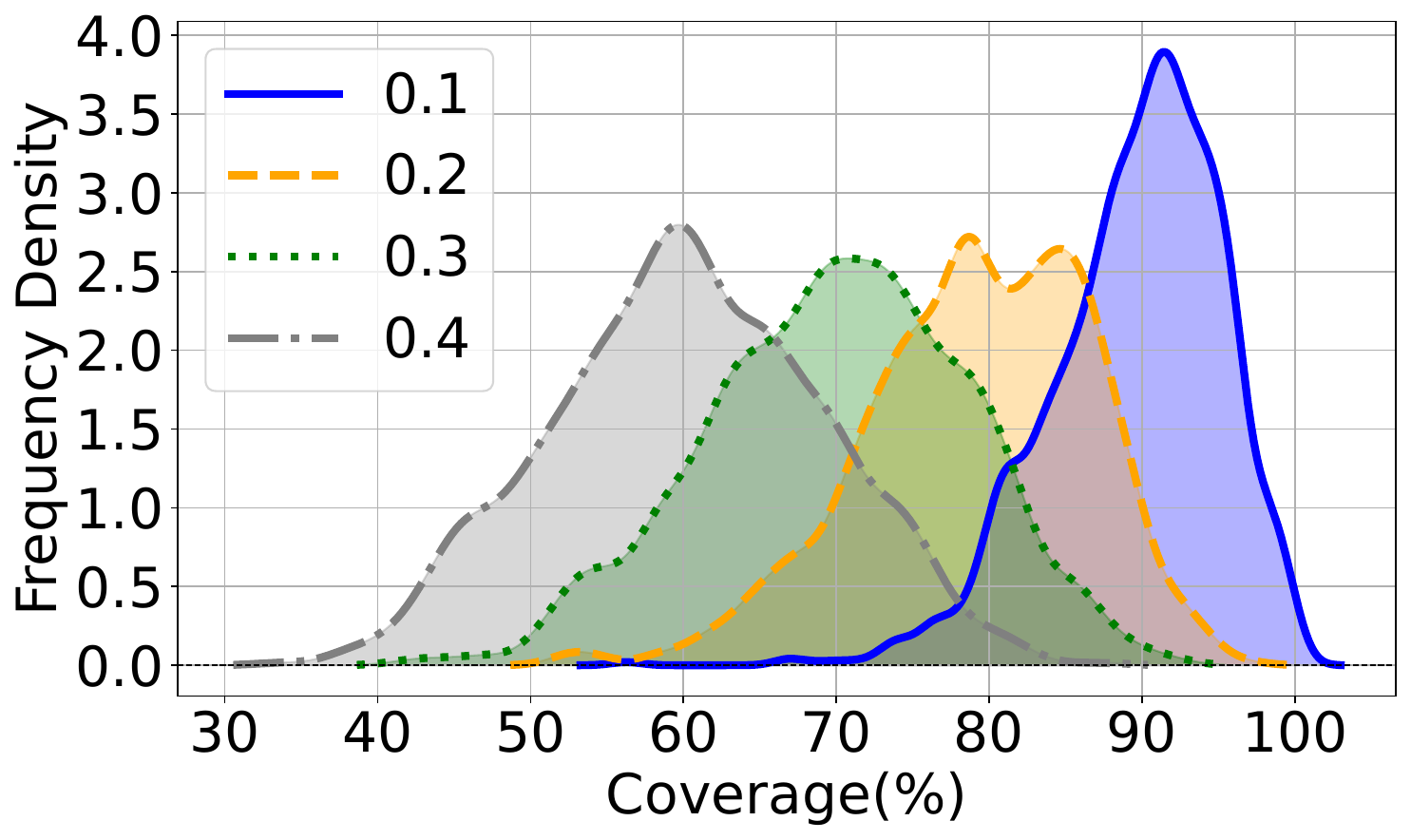}
        \vspace{-8pt}\caption{Balsa on JOB}
        \label{fig:deltaExperimentsIMDB}
    \end{subfigure}
    % \hfill
    \begin{subfigure}{0.48\linewidth}
        \centering
        \includegraphics[width=0.8\linewidth]{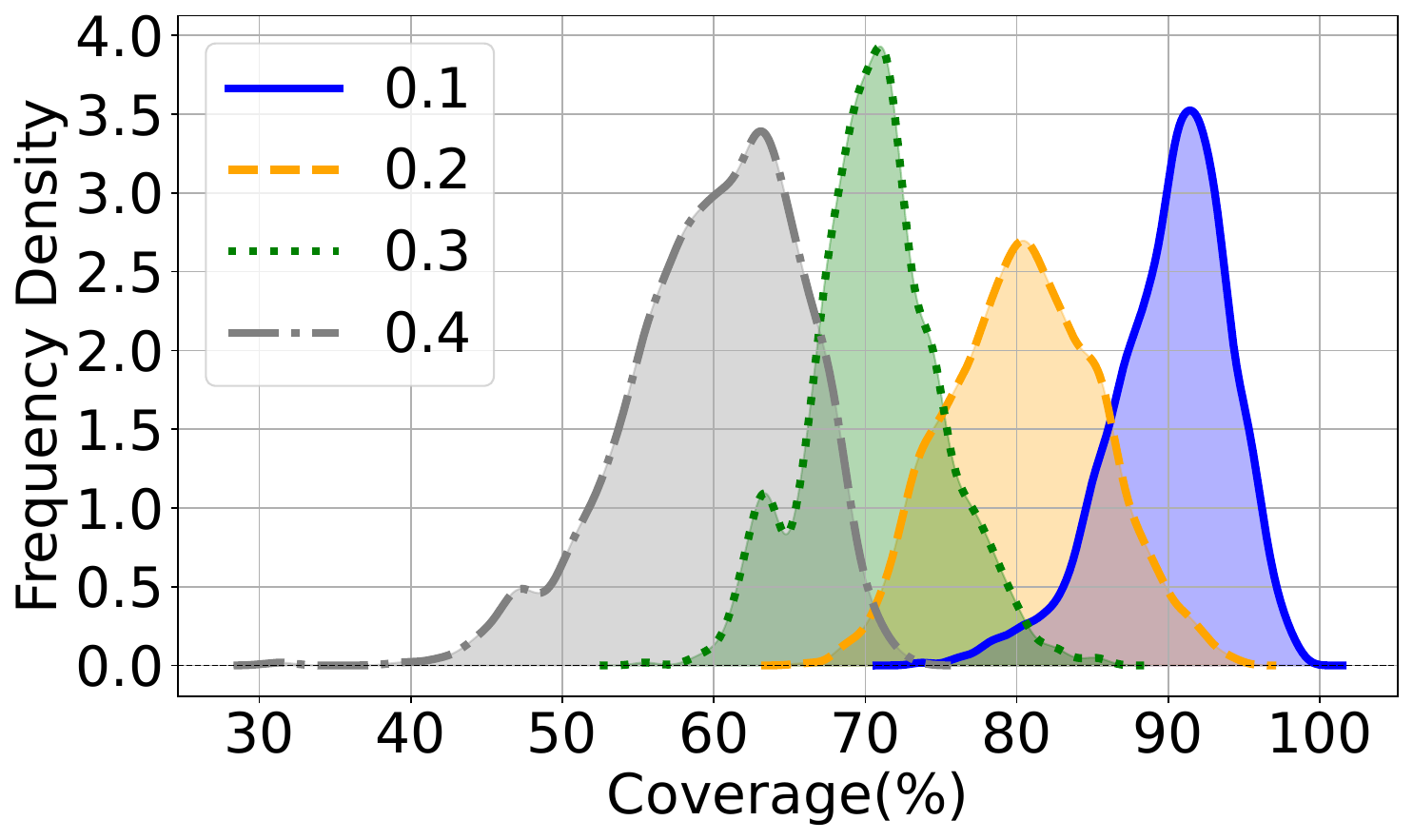}
        \vspace{-8pt}\caption{Balsa on TPC-H}
        \label{fig:deltaExperimentsTPCH}
    \end{subfigure}
    
    % \vspace{0.3cm} 
    
    \begin{subfigure}{0.48\linewidth}
        \centering
        \includegraphics[width=0.8\linewidth]{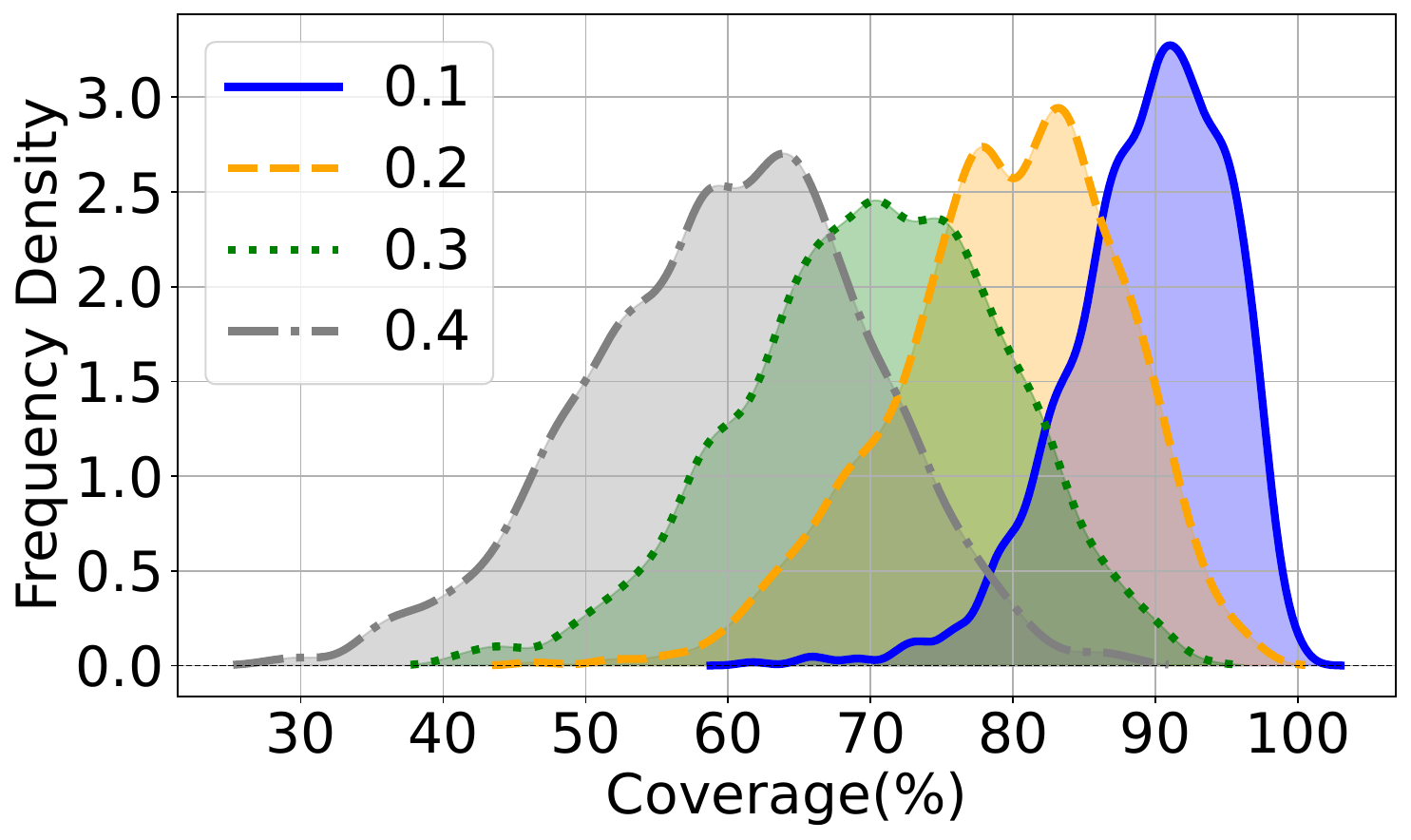}
        \vspace{-8pt}\caption{Lero on JOB}
        \label{fig:leroExperimentsIMDB}
    \end{subfigure}
    % \hfill
    \begin{subfigure}{0.48\linewidth}
        \centering
        \includegraphics[width=0.8\linewidth]{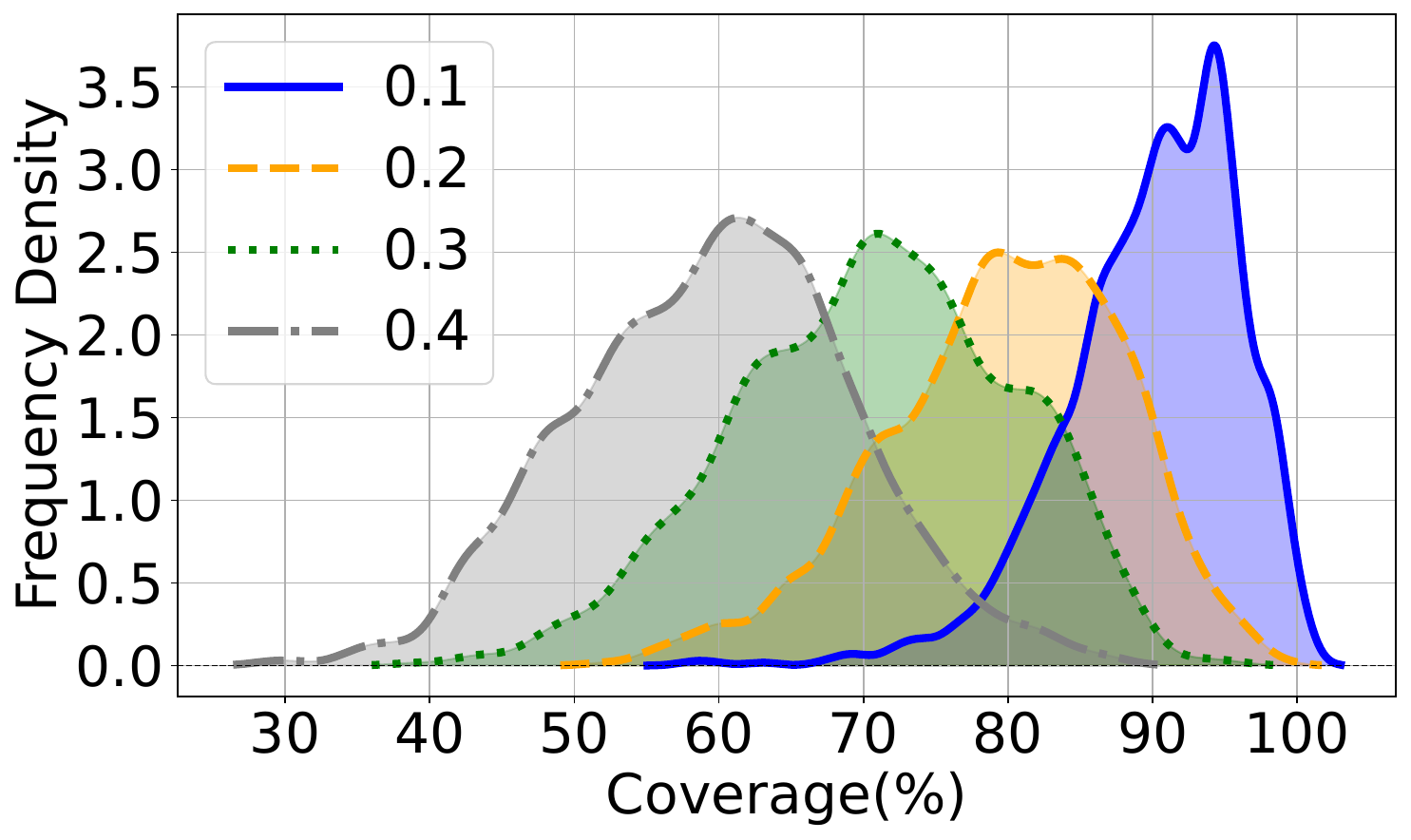}
        \vspace{-8pt}\caption{Lero on TPC-H}
        \label{fig:leroExperimentsTPCH}
    \end{subfigure}
\vspace{-8pt}
\caption{Impact on Choice of $\delta$.}
\label{fig:impact-delta}
\vspace{-10pt}\end{figure}

\noindent{\textbf{Impact of Uncertainty Probability $\delta$.}}
The second hyper-parameter is the uncertainty probability $\delta$. We varied $\delta$ across four values: 0.1, 0.2, 0.3, and 0.4. Similar to the previous discussion, we expect the peaks of the coverage curve to align with $1 - \delta$, meaning the corresponding peaks should align with 0.9, 0.8, 0.7, and 0.6. Figure~\ref{fig:impact-delta} illustrates this trend. Additionally, we observe that as $\delta$ decreases, the area under the curve becomes sharper and narrower, indicating a more concentrated coverage distribution. This suggests that with smaller values of $\delta$ (e.g., $\delta=0.1$ in our graph), obtaining $C$ values in a single sampling iteration is more likely to yield values centered around the expected confidence level of $1 - \delta$. %Based on this observation, we set $\delta = 0.1$ as the default for the following experiments.
%This experiment demonstrates the flexibility and robustness of our {\cp}-based latency guarantee method in handling various user-defined confidence levels.

% Additionally, we observe that as $\delta$ increases,  the area under the curve becomes narrower, indicating that coverage distribution is more concentrated. This suggests that for each sampling iteration (actual application), the $C$ values we obtain are more reliably centered around the expected confidence $1 - \delta$. Based on this observation, we will use $\delta = 0.1$ as the default in the following experiments.

% Old graph - Oct29
% \begin{figure}[h]
%     \centering
%     \includegraphics[width=1\linewidth]{figures//1000iter//BalsaXImdb/1-Delta}
%     \caption{Impact on choice of $\delta$ - [Balsa x JOB]}
%     \label{fig:deltaExperiments}
% \end{figure}

% \begin{figure}[h]
%     \centering
%     \includegraphics[width=1\linewidth]{figures//1000iter//BalsaXImdb/BalsaXTPCHDelta.png}
%     \caption{Impact on choice of $\delta$ - [Balsa x TPCH]}
%     \label{fig:enter-label}
% \end{figure}

% \begin{figure}[h]
%     \centering
%     \includegraphics[width=1\linewidth]{figures//1000iter//BalsaXImdb/LeroXIMBD.png}
%     \caption{Impact on choice of $\delta$ - [Lero x JOB]}
%     \label{fig:enter-label}
% \end{figure}

% \begin{figure}
%     \centering
%     \includegraphics[width=1\linewidth]{figures//1000iter//BalsaXImdb/leroXtpchDelta.png}
%     \caption{Impact on choice of $\delta$ - [Lero x TPCH]}
%     \label{fig:enter-label}
% \end{figure}

\vspace{-9pt}\section{Related Work}

\textbf{Learned Query Optimization ({\lqo}).} In recent years, numerous {\ml}-based techniques have been proposed to improve query optimization. One direction is to use {\ml} to improve cardinality estimates for query outputs and use them to predict query plan costs~\cite{deep_card_est2, neurocard, unsupervised_card19, learnedce21,learned_cost_estimator_1,learned_cost_estimator_2,flowloss}. Although this direction has shown improved cardinality estimation accuracy, it does not provide evidence that such improvements result in better query plans~\cite{flowloss}.
Consequently, two lines of work have emerged to directly learn how to optimize the query plan itself (e.g.,~\cite{neo, bao, balsa,kepler, zhu2023lero, yu2020reinforcement,marcus2018deep,krishnan2018learning,yu2020reinforcement}), either by constructing the plan from scratch (e.g.,~\cite{neo, balsa}) or by choosing among different candidate plans generated by traditional optimizers (e.g.,~\cite{zhu2023lero,bao,kepler,yu2020reinforcement}). 
Examples of the first line of work include {\neo}~\cite{neo} and {\balsa}~\cite{balsa}. {\neo} introduces a novel query encoding technique, namely Tree Convolution, to capture execution patterns in the plan tree. In contrast, {\balsa} reduces reliance on experts by employing a simulation-to-reality learning framework. Examples of the second line of work include {\bao}~\cite{bao} and {\lero}~\cite{zhu2023lero}.  {\bao}~\cite{bao} employs a multi-armed bandit approach to estimate the costs of candidate plans generated by the traditional optimizer and select the best among them. {\lero}~\cite{zhu2023lero} takes a unique approach by constructing a learned model that performs pairwise plan comparisons rather than traditional cost or latency prediction. 

Although all {\lqo}s have demonstrated improved query performance, they typically do not consider the robustness issues (no guarantees on stability or regression avoidance). \revised{{\kepler}~\cite{kepler} and Roq~\cite{roq} are the closest works to our objective. Kepler employs robust neural network prediction techniques to reduce tail latency and minimize query regressions. Specifically, it utilizes Spectral-normalized Neural Gaussian Processes~\cite{sngp20} to quantify its confidence in plan prediction and falls back to the traditional optimizer when uncertain. Roq introduces robustness notions in the context of query optimization and incorporates a complex {\ml} pipeline to predict plan cost and risk. However, neither method provides theoretical guarantees or formally formulates {\lqo} plan construction verification.}
% {\kepler}~\cite{kepler} is the closest {\lqo} to our objective in this paper. It uses robust neural network prediction techniques to decrease tail latency and minimize query regressions (i.e., avoiding scenarios where the performance of predictions is worse than the traditional optimizer). Specifically, it employs Spectral-normalized Neural Gaussian Processes~\cite{sngp20} to quantify its confidence about a plan prediction and falls back to the traditional optimizer when uncertain. However, {\kepler} provides no theoretical or formal guarantees on predicted query plans.
To our knowledge, our work is the first to address the verification problem in {\lqo}s by providing formal guarantees and using them to guide the plan construction process.

\noindent\textbf{Conformal Prediction ({\cp}).} {\cp} was originally introduced to provide a robust statistical framework for quantifying prediction uncertainty (e.g.,~\cite{cpbook05,shafer2007tutorialconformalprediction,cpgentle23}). Extensive research has explored the application of {\cp} in distribution-agnostic settings, delivering reliable performance guarantees even in non-stationary environments (e.g.,~\cite{angelopoulos2022gentleintroductionconformalprediction, fontana2023conformal, lei2018distribution, zhao2024robust, prtstl22}). Additionally, extensions of {\cp} have been applied to time-series data~\cite{cpcovshift19, cauchois2020robust} and ({\stl})-based runtime verification in real-time systems (e.g., autonomous cars~\cite{cpstl23}, autonomous robots~\cite{prtrobots23}, aircraft simulation~\cite{cpstl23,prtstl22}).  \revised{Recently, several work discuss applying {\cp} within different distribution shift conditions~\cite{stf-adptive-cp, cp-covariate-shift, cp-beyond-exchangeability, zhao2024robust}.} Besides, {\cp} has been adapted for policy evaluation in reinforcement learning~\cite{foffano2023conformaloffpolicyevaluationmarkov, taufiq2022conformal}, time-series forecasting~\cite{NEURIPS2021_312f1ba2}, and outlier detection~\cite{Bates_2023}. It has also been employed to monitor risks in evolving data streams~\cite{podkopaev2022trackingriskdeployedmodel} and detect change points in time-series data~\cite{Vovk_2021, pmlr-v60-volkhonskiy17a}. 

% These diverse applications underscore the flexibility and significance of {\cp} in ensuring reliable performance across various domains. 

% \textbf{Signal Temporal Logic.} Conformal prediction was used for verification of STL properties in by learning a predictive model of the STL semantics. For reachable set prediction, the authors in [69–71] used conformal prediction to quantify uncertainty of a predictive runtime monitor that predicts reachability of safe/unsafe states. However, the works in [68–71] train task-specific predictors while we use task-independent trajectory predictors to predict future system states from which we infer information about the satisfaction of the task. This is significant as no expensive retraining is required when the specification changes.

%\input{sections/related-work}
% \section{Future Extension}

% Generic -> Query template to quantify the difficulties.
\vspace{-2pt}% \section{Conclusion}
% In conclusion, our framework represents a pioneering application of conformal prediction to the verification of learned query optimizers, addressing a critical need for robustness and reliability. We have demonstrated that conformal prediction can offer a flexible, lightweight verification method that ensures trustworthy performance boundaries while maintaining efficient query processing. By providing both runtime and offline verification modes, our approach enables deployment in real-world production environments, achieving formal verification without incurring significant computational overhead. The high degree of accuracy in our experimental results across multiple optimizers, including Balsa, Lero, and TreeLSTM, underscores the potential for conformal prediction to become a foundational component in learned database systems. Our framework not only validates learned query optimizers but also lays the groundwork for broader applications of statistical formal verification across learned components in database management systems.
\vspace{-0.25pt}
\section{Conclusion}
To the best of our knowledge, we are the first to introduce Conformal Predication ({\cp}) for verifying learned database components, with a focus on the learned query optimization ({\lqo}). Our framework encompasses {\cp}-based latency \revised{bounds} across multiple granularities and runtime verification. \revised{Our framework also employs an adaptive {\cp} approach for handling distribution shifts.} Further, we introduce a {\cp}-guided query optimization algorithm capable of enhancing {\lqo}s. We have demonstrated that {\cp} provides a flexible, lightweight verification approach that establishes trustworthy prediction boundaries. Our methods can be deployed in real-world production environments, achieving formal verification without significant computational overhead. Our evaluation shows that {\cp} can be used to achieve tight upper bounds on actual latency using predicted cost. \revised{Adaptive $\cp$ maintains the confidence levels even under distribution shift}. CP-guided LQOs produce plans with up to 9x better actual latency, over the entire workload CP-guided LQOs show a 9.96\% reduction in actual latency. Our {\cp} framework also lays the groundwork for broader applications across various learned components within database systems.

% improves both query plan
% quality (up to 9.84x) and planning time, with a 74.4% reduction for
% Query 4b and a 9.96% reduction across all test queries from trained
% LQOs

% The comprehensive experimental evaluations across multiple optimizers, including Balsa, Lero, and RTOS, prove the potential of {\cp} application. 

% The comprehensive experimental evaluations across multiple optimizers, including Balsa, Lero, and RTOS, highlight the potential for conformal prediction to serve as a foundational component in learned database systems. Our {\cp}  framework not only supports {\lqo}s but also establishes a foundation for broader applications across other learned components in database systems.

%\newpage

\section{Acknowledgments}
We sincerely thank our USC colleagues Yiqi Zhao, Lars Lindemann, and Jyotirmoy V. Deshmukh for their valuable insights, constructive feedback, and continuous support throughout the development of this work.

%\begin{acks}
%To Robert, for the bagels and explaining CMYK and color spaces.
%\end{acks}

\bibliographystyle{ACM-Reference-Format}
\bibliography{sample-base,cp_learned_db}

\appendix

\end{document}